\documentclass{article}

\usepackage[top=3cm, bottom=3cm, left=3cm,right=3cm]{geometry}
\usepackage[colorinlistoftodos]{todonotes}
\usepackage{graphicx}
\usepackage{amssymb}
\usepackage{amsmath}
\usepackage{gensymb}
\usepackage{bbm}
\usepackage{todonotes}
\usepackage{pdflscape}
\usepackage{caption}
\usepackage{subcaption}
\usepackage[square,numbers]{natbib}
\usepackage[T1]{fontenc}
\usepackage[utf8]{inputenc}
\usepackage{authblk}
\usepackage{gensymb}
\usepackage{pdfpages}
\usepackage{setspace} 
\usepackage{booktabs}
\usepackage{longtable}
\usepackage{float}
\usepackage{tikz}
\usepackage[colorlinks=true,citecolor=blue, linkcolor=blue]{hyperref}
\usepackage{multirow}
\usepackage[parfill]{parskip}

\usepackage{tikz}
\usetikzlibrary{shapes.geometric, arrows}

\tikzstyle{startstop} = [rectangle, rounded corners, minimum width=3cm, minimum height=1.75cm, text centered, draw=black, text width=3cm, fill=red!30]
\tikzstyle{bigrec} = [rectangle, rounded corners, minimum width=8.25cm, minimum height=1.75cm, text centered, draw=black, text width=5cm, fill=blue!30]
\tikzstyle{bigrec2} = [rectangle, rounded corners, minimum width=11.2cm, minimum height=1.75cm, text centered, draw=black, text width=3cm, fill=green!30]

\tikzstyle{arrow} = [thick, ->, >=stealth]
\tikzstyle{darrow} = [thick, <->, >=stealth]

\title{A multi-level model for estimating region-age-time-type specific male circumcision coverage from household survey and health system data in South Africa}

\author[1,2]{Matthew L. Thomas}
\author[3,4]{Khangelani Zuma}
\author[5]{Dayanund Loykissoonlal}
\author[6]{Bridget Dube}
\author[7]{Peter Vranken}
\author[7]{Sarah E. Porter}
\author[8]{Katharine Kripke}
\author[5,9]{Thapelo Seatlhodi}
\author[10,11,12]{Gesine Meyer-Rath}
\author[9]{Leigh F. Johnson}
\author[2]{Jeffrey W. Eaton}

\affil[1]{Joint Centre for Excellence in Environmental Intelligence, University of Exeter \& Met Office, Exeter, United Kingdom}
\affil[2]{MRC Centre for Global Infectious Disease Analysis, School of Public Health, Imperial College London, London, United Kingdom}
\affil[3]{Human and Social Capabilities Research Division, Human Sciences Research Council,  Pretoria, South Africa}
\affil[4]{School of Public Health, University of the Witwatersrand, Johannesburg, South Africa}
\affil[5]{National Department of Health, Pretoria, South Africa}
\affil[6]{Genesis Analytics, Johannesburg, South Africa}
\affil[7]{Division of Global HIV and Tuberculosis, Centers for Disease Control and Prevention, Pretoria, South Africa}
\affil[8]{Avenir Health, Washington, United States of America}
\affil[9]{Centre for Infectious Disease Epidemiology and Research, University of Cape Town, Cape Town, South Africa}
\affil[10]{Health Economics and Epidemiology Research Office, University of Witwatersrand, Johannesburg, South Africa}
\affil[11]{Department of Medicine, Faculty of Health Sciences, University of Witwatersrand, Johannesburg, South Africa}
\affil[12]{Department of Global Health, Boston University School of Public Health, Boston, USA}

\begin{document}


\maketitle
\newpage
\begin{abstract}
	\noindent Voluntary medical male circumcision (VMMC) reduces the risk of male HIV acquisition by 60\%. Programmes to scale-up VMMC to prevent HIV infection have been introduced in sub-Saharan African countries with high HIV burden. While large-scale provision of VMMC is recent, traditional male circumcision (MC) has long been conducted as part of male coming-of-age practices. How and at what age traditional MC occurs varies by ethnic groups and geographies. Accurate estimates of MC coverage by age and type of circumcision (traditional or medical) over time at sub-national levels are essential for planning and delivering VMMCs to meet targets and evaluating their impacts on HIV incidence. In this paper, we developed a Bayesian competing risks time-to-event model to produce region-age-time-type specific probabilities and coverage of MC with probabilistic uncertainty. The model jointly synthesises data from household surveys and health system data on the number of VMMCs conducted. We demonstrated the model using data from five household surveys and VMMC programme data to produce estimates of MC coverage for 52 districts in South Africa between 2008 and 2019. Nationally in 2008, 24.1\% (CI: 23.4-24.8\%) of men aged 15-49 were traditionally circumcised and 19.4\% (CI: 18.9-20.0\%) were medically circumcised. Between 2010 and 2019, 4.25 million VMMCs were conducted, and MC coverage among men aged 15-49 increased to 64.0\% (CI: 63.2-64.9\%) and medical MC coverage to 42\% (CI: 41.3-43.0\%). MC coverage varied widely across districts, ranging from 13.4-86.3\%. The average age of traditional MC ranged between 13 to 19 years, depending on local cultural practices.
\end{abstract}

\newpage 


\section{Introduction}


Preventing HIV continues to be a major public health priority, particularly in southern and eastern Africa. Despite considerable progress combating the epidemic, the numbers of new HIV infections remain well above national and international targets \citep{UNAIDSHighLevel}. Voluntary medical male circumcision (VMMC), defined as the complete surgical removal of the foreskin, has emerged as an effective intervention to reduce the risk of HIV acquisition among men. It is estimated to reduce the risk of female-to-male transition of HIV by 60\% \citep{gray2007male, bailey2007male, auvert2005randomized, gray2012effectiveness, grund2017association}. There is some evidence that VMMC may also reduce the risk of HIV acquisition among men who have sex with men \citep{pintye2019benefits} and may decrease the risk of other sexually transmitted infections, such as syphilis, herpes simplex virus type 2, and human papillomavirus \citep{tobian2009male}. 

VMMC is appealing as an HIV prevention intervention because it is a one-time, efficient, safe, and cost-effective method. The World Health Organization (WHO) and the Joint United Nations Programme on HIV/AIDS (UNAIDS) have identified fifteen African countries with high HIV prevalence and low male circumcision (MC) prevalence as priority countries to scale-up VMMC for HIV prevention \citep{UNAIDSJoint, davis2018progress, WHOVoluntary2}. In 2010, ambitious targets were set to achieve 80\% circumcision coverage in men aged 15--49 years by 2015, and while countries are still working towards meeting this, new complementary targets of 90\% circumcision coverage in adolescent boys and young men aged 10--29 years by 2021 were set in 2016  \citep{WHOFramework}. In South Africa, the \textit{South African National Strategic Plan for HIV, TB, and STIs 2017-2022} set a target to provide 2.5 million VMMC over the five years as part of a comprehensive combination HIV prevention package \cite{sanac}.

While large-scale provision of medical male circumcision (MMC) for HIV prevention is relatively recent, male circumcision (MC) has traditionally been practiced in many African countries as part of traditional male initiation ceremonies (TMIC). Circumcisions during TMIC are typically conducted among adolescent boys and young men, but its provision, typical age at circumcision, and what it entails differ considerably across and within countries influenced by community-established values, religious, ethnic and cultural identities. In South Africa, there is substantial heterogeneity in traditional circumcision practices across ethnic groups, ranging from rarely conducted to nearly universal \citep{peltzer2014prevalence, connolly2008male}. Traditional male circumcision (TMC) conducted during TMIC is often performed using non-medical methods in non-clinical settings by a traditional practitioner typically with no formal medical training \citep{drain2006male, wilcken2010traditional, weiss2000male, wilcken2010traditional}. Whether TMC involves complete removal of the foreskin varies by practicing groups. In some cases, TMC may involve partial circumcision or a simple incision in the prepuce and is not thought to confer the same HIV prevention benefits as medical circumcision \citep{WHOTraditional, shaffer2007protective, bailey2008male}. Due to this, VMMC programmes are increasingly working with traditional leaders for MMCs to be conducted as part of TMICs by medical service providers (hereafter referred to as MMC-T for medical male circumcisions conducted in traditional settings). Replacing TMC with MMC-T ensures men circumcised through TMIC have the same protection against HIV infection, as well as safe and sanitary surgical procedures.  \citep{WHOTraditional}. 

In light of this heterogeneity, planning and delivering VMMC services to meet programmatic targets requires detailed information about the coverage of circumcision by age group and type of circumcision (traditional or medical) at subnational level. Data about the MC coverage is available from two sources: (i)   household surveys collecting self-reported circumcision status, and (ii) health system programme data on the number of VMMC conducted for HIV prevention. National household surveys, conducted roughly every five years, collect data on current HIV status and other risk factors including the self-reported circumcision status, age at circumcision, and type of circumcision from nationally-representative samples. Household survey estimates of MC coverage are relatively precise at the national level, but have large sampling variation at sub-national levels and granular age groups due to small sample sizes. The numbers of VMMC conducted by HIV prevention services are reported to the National Department of Health (NDoH) by VMMC programme implementers, but this does not reflect men circumcised traditionally or medical circumcisions conducted before the start of the national VMMC programme in 2010.

Previous approaches to estimate district-level circumcision coverage have considered both data sources, but often not together. Cork \textit{et al.} analysed household survey data using a Bayesian geo-statistical model producing annual estimates of total MC coverage in men aged 15-49 years in sub-Saharan Africa at multiple spatial resolutions between 2000 and 2017 \citep{cork2020mapping}. The \textit{Decision Makers' Program Planning Toolkit, Version 2} (DMPPT2) is a model to support planning VMMC scale-up in sub-Saharan Africa to allow countries to enter reported number of VMMCs as well as generate VMMC targets, coverage estimates, and impact projections. DMPPT2 combines estimates of MC coverage prior to MC scale-up (estimated using household surveys) with the reported numbers of VMMCs from various programmes to estimate MC coverage and unmet need at a subnational level over time. This approach is limited as it does not incorporate data from more recent household surveys and does not provide the statistical uncertainty associated with the MC coverage estimates \citep{kripke2016age, kripke2016cost}.

We developed a model that synthesises both survey and VMMC programme data to estimate the probabilities and coverage of MMC and TMC by space, age, and time with probabilistic uncertainty. Set within a Bayesian hierarchical framework, the model comprises of two components; (i) a time-to-event analysis in which survey data is used to determine space-age-time-type specific probabilities and coverage of MC and (ii) a Poisson modelling for the number of VMMC conducted for HIV prevention to further inform probability of MMC. The remainder of the paper is organised as follows: Section 2 describes the statistical modelling framework used to estimate space-age-time-type specific probabilities and coverage of MC. In Section 3, the model is implemented to estimate annual probabilities and coverage of circumcision using data from South Africa. Finally, Section 4 provides a concluding summary and a discussion of potential areas for future research.


\section{Statistical Methods}
\label{sec::methods}


Using a Bayesian hierarchical model with small area estimation methods, we modelled probabilities of circumcision stratified by region, age, and time for two types of circumcision: (1) circumcisions that occurred in traditional male initiation ceremonies or for other religious or cultural reasons (TMIC) and (2) circumcisions for non-traditional reasons and/or HIV prevention that take place in a clinical setting using medical methods (MMC-nT). Reflecting recent efforts to encourage adoption of medical methods in TMICs, TMICs were sub-categorised into: (1) traditional circumcisions conducted using non-medical methods (TMC) and (2) circumcisions conducted as part of TMIC but using medical methods (MMC-T), determined by a district- and year-specific probability that TMIC circumcisions were performed as MMC-Ts.

Probabilities of circumcision were estimated using a competing risk discrete time-to-event model \citep{putter2006tutorial}, with estimates of circumcision coverage by type within each cohort calculated using the cumulative incidence. These were combined with data on male population size to calculate the predicted number of circumcisions conducted per year. Likelihood functions were specified for the two data sources to inform parameter calibration: (1) the probability of individual-level observations of circumcision age, type, and year reported by men in national household surveys in a time-to-event framework, and (2) the reported number of medical male circumcisions conducted in each district for HIV prevention among males 10 years and older using a Poisson count model. 

\begin{figure}[H]
	\centering
	\begin{tikzpicture}[node distance=2cm]
		\draw [black, fill=gray!30] (-9cm,-1.75cm) -- (6cm,-1.75cm) -- (6cm,-7.25cm) -- (-9cm,-7.25cm) -- cycle;
		\node (MC) [bigrec2] {Survey data};
		\node (MMC) [startstop, below of = MC, right of = MC, xshift = 2cm, yshift = -1cm] {MMC-nT \\ $\lambda^{\text{MMC-nT}}$};
		\node (TMIC) [startstop, below of = MC, left of = MC, xshift = -2cm, yshift = -1cm] {TMIC \\ $\lambda^{\text{TMIC}}$};
		\node (TMC) [startstop, below of = TMIC, left of = TMIC, xshift = -1cm, yshift = -1cm] {TMC \\ $(1-p)\cdot\lambda^{\text{TMIC}}$};
		\node (MMCT) [startstop, below of = TMIC, right of = TMIC, xshift = 1cm, yshift = -1cm] {MMC-T \\ $p\cdot\lambda^{\text{TMIC}}$};
		\node (ProgData) [bigrec, below of = MC, xshift = 1.51cm, yshift = -7cm] {VMMC programme data};
		\draw [arrow] (MC.south -| TMIC.north) |- ++(0,-5mm) -| (TMIC.north);
		\draw [arrow] (MC.south -| MMC.north) |- ++(0,-5mm) -| (MMC.north);
		\draw [arrow] (TMIC.south) |- ++(0,-5mm) -| (MMCT.north);
		\draw [arrow] (TMIC.south) |- ++(0,-5mm) -| (TMC.north);
		\draw [arrow] (ProgData.north -| MMC.north) -- (MMC.south);
		\draw [arrow] (ProgData.north -| MMCT.north) -- (MMCT.south);
	\end{tikzpicture}		
	\caption{Schematic representation of types of circumcision included in the model and data sources informing each. TMIC = circumcisions conducted as part of traditional male initiation ceremonies; TMC = traditional male circumcisions (conducted by a non-medical practitioner); MMC-T = medical male circumcision conducted as part of traditional male initiation practices; MMC-nT = medical male circumcision conducted outside the context of traditional male initiation (the majority of VMMC for HIV prevention and other routine medical male circumcision).}
	\label{fig::modelstr}
\end{figure}
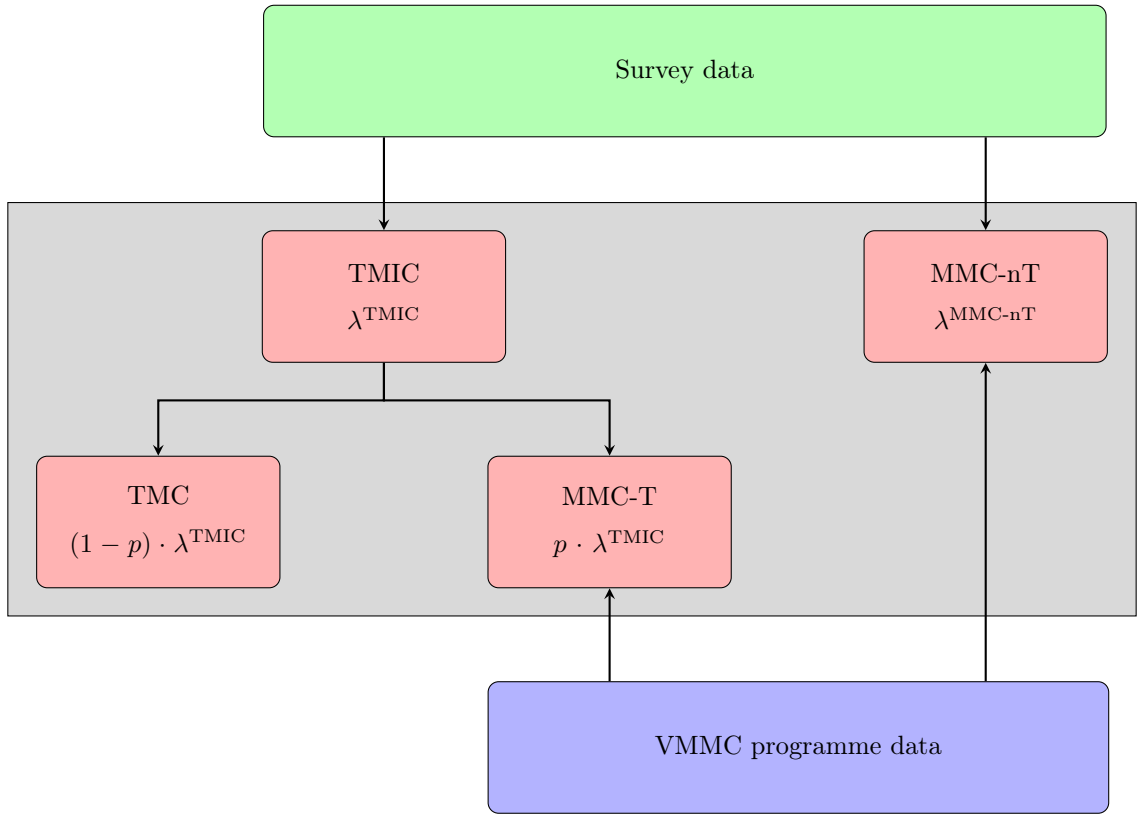

Throughout the methods and applications sections, we refer to the following types of circumcision (Figure \ref{fig::modelstr}):
\begin{itemize}
    \item \textbf{MMC-nT:} Medical male circumcisions conducted outside of traditional male initiation ceremonies, representing the large majority of MMC conducted.
    \item \textbf{TMC:} Traditional male circumcisions, assumed to be conducted outside a medical setting for traditional male initiation purposes.
    \item \textbf{MMC-T:} Medical male circumcisions conducted as part of traditional male initiation ceremonies, typically in place of circumcisions that previously would have been conducted as TMC.
\end{itemize}
Useful aggregates of these circumcision types are referred to as:
\begin{itemize}
    \item \textbf{MMC:} All medical male circumcisions (MMC-nT + MMC-T), assumed to be consistent with circumcisions reported through VMMC programme data reporting.
    \item \textbf{TMIC:} All male circumcisions conducted as part of traditional male initiation practices (TMC + MMC-T).
    \item \textbf{MC:} All male circumcisions of any type (MMC-nT + TMC + MMC-T).
\end{itemize}

\subsection{Process model}
\label{sec::competingrisks}


\subsubsection{Probabilities of circumcision per time step}


We modelled the probabilities of becoming circumcised per time step through either TMIC or MMC-nT for residents of region $i \in I = \{1, 2, \ldots, N_I\}$ having an MMC, MMC-nT, MMC-T, TMC, TMIC or MMC at age $a \in A = \{0, 1, 2, \ldots, N_A\}$ and time $t \in T = \{1, 2, \ldots, N_T\}$.

We defined $\lambda^{\text{TMIC}}_{iat}$ as the probability that an individual in region $i$ received TMIC at age $a$ and time step $t$, given they were uncircumcised by age $a-1$ and time $t-1$, 
\begin{equation} 
		\lambda^{\text{TMIC}}_{iat} = \mathbb{P}(\text{TMIC} \; \text{in} \; (i,a,t) \; | \; \text{Uncircumcised in} \; (i,a-1, t-1)). 
	\label{eqn::tmic}
\end{equation}
This was modelled using piece-wise logit-linear function:
\begin{equation*} 
	\text{logit}(\lambda^{\text{TMIC}}_{iat}) = \hat{\alpha} + \hat{\psi_i} + \hat{\phi}_a + \hat{\gamma}_{ia}
\end{equation*}
where $\hat{\alpha}$ is the intercept, $\hat{\psi_i}$ is a regional random effect, $\hat{\phi}_a$ is an age random effect, and $\hat{\gamma}_{ia}$ is an age-space interaction term to allow different age patterns of TMIC  across regions. We assumed that the probability of TMIC was constant over time (i.e. $\lambda^{\text{TMIC}}_{iat} \equiv \lambda^{\text{TMIC}}_{ia}$) as TMIC practices have been relatively stable over time. 

Similarly, we defined $\lambda^{\text{MMC-nT}}_{iat}$ as the probability an individual in region $i$ received MMC-nT at age $a$ and time $t$, $\lambda^{\text{MMC-T}}_{iat}$, given they were uncircumcised by age $a-1$ and time $t-1$,  
\begin{equation}
		\lambda^{\text{MMC-nT}}_{iat} = \mathbb{P}(\text{MMC-nT} \; \text{in} \; (i,a,t) \; | \; \text{Uncircumcised in} \; (i,a-1, t-1)),
	\label{eqn::mmc}
\end{equation}
To ensure the overall probability of circumcision $\lambda^{\text{TMIC}}_{iat} + \lambda^{\text{MMC-nT}}_{iat} \leq 1$ for all $i$, $a$ and $t$ in the discrete-time framework, we applied the probability of TMIC  before MMC-nT in each time step, and therefore initially model the probability an individual in region $i$ received MMC-nT at age $a$ and time $t$, $\lambda^{\text{MMC-T}}_{iat}$, given they were uncircumcised by age $a-1$ and time $t-1$ and did not recieve TMIC at age $a$ and time $t$
\begin{equation*}
		\tilde{\lambda}^{\text{MMC-nT}}_{iat} = \mathbb{P}(\text{MMC-nT} \; \text{in} \; (i,a,t) \; | \; \text{Uncircumcised in} \; (i,a-1, t-1) \text{ and no TMIC in} \; (i,a, t))
\end{equation*}
such that
\begin{equation*}
		\lambda^{\text{MMC-nT}}_{iat} = \tilde{\lambda}^{\text{MMC-nT}}_{iat}\cdot (1-\lambda^{\text{TMIC}}_{iat}).
\end{equation*}
The probability $\tilde{\lambda}^{\text{MMC-nT}}_{iat}$ was separated into two processes: (i) paediatric circumcision (for those aged 0--9) and (ii) adolescent and adult circumcision (for those aged 10 and over). VMMC programmes only provide MMCs for HIV prevention to those aged 10 and over, while infant and paediatric medical circumcision tends to occur through cultural or religious practices unrelated to scale-up of VMMC for HIV prevention. Therefore, we modelled $\tilde{\lambda}^{\text{MMC-nT}}_{iat}$ using a piece-wise logit-linear function:
\begin{equation*}
	\text{logit}(\tilde{\lambda}^{\text{MMC-nT}}_{iat}) =
	\begin{cases}
		\bar{\alpha} + \bar{\psi}_i + \bar{\phi}_a + \bar{\gamma}_{ia} & \text{for } 0 \leq a \leq 9\\
		\alpha + \psi_i + \phi_a + \theta_t + \gamma_{ia} + \delta_{at} + \zeta_{it} & \text{for } a \geq 10
	\end{cases} 
\end{equation*}
where $\bar{\alpha}$ and $\alpha$ are intercepts, $\bar{\psi}_i$ and $\psi_i$ and  are regional random effects, $\bar{\phi}_a$ and $\phi_a$ are age random effects, $\bar{\gamma}_{ia}$ and $\gamma_{ia}$ are age-space interaction terms to allow for different age patterns of MMC-nTs between regions, for paediatric circumcision and adolescent and adult circumcision respectively. The $\delta_{at}$ is an age-space interaction term to allow for different age patterns of MMC-nTs over time and $\zeta_{it}$ is an space-time interaction term to allow for different uptake of MMC-nTs over time across regions, in adolescent and adult men. Similarly, we assumed that the probabilities of MMC-nT among those aged 0--9 were constant over time, (i.e. $\lambda^{\text{MMC-nT}}_{iat} \equiv \lambda^{\text{MMC-nT}}_{ia}$ when $0\leq a \leq 9$), as paediatric medical circumcision practices are unrelated to VMMC for HIV prevention and are relatively stable. 

For individuals circumcised in TMIC, we defined a probability, $p_{iat}$ that circumcision was conducted as a MMC-T versus TMC, specified for region $i$, age $a$, and time $t$. The probability an individual in region $i$ received MMC-T at age $a$ and time $t$, $\lambda^{\text{MMC-T}}_{iat}$ was thus 
\begin{equation}
	\begin{split}
		\lambda^{\text{MMC-T}}_{iat} &= \mathbb{P}(\text{MMC-T} \; \text{in} \; (i,a,t) \; | \; \text{Uncircumcised in} \; (i,a-1, t-1)) \\
		&=p_{iat}\cdot\lambda^{\text{TMIC}}_{iat}
	\end{split}
	\label{eqn::mmct}
\end{equation}
and  the probability an individual in region $i$ received TMIC was
\begin{equation}
	\begin{split}
		\lambda^{\text{TMC}}_{iat} &= \mathbb{P}(\text{TMC} \; \text{in} \; (i,a,t) \; | \; \text{Uncircumcised in} \; (i,a-1, t-1)) \\
		&=(1-p_{iat})\cdot\lambda^{\text{TMIC}}_{iat}
	\end{split}
	\label{eqn::mmct}
\end{equation}
Prior to VMMC programmes, all circumcisions conducted in TMIC were TMC, in which case $p_{iat}=0$. The proportion of TMICs conducted as MMC-Ts may not be identifiable from the survey data, particularly for years since the most recent survey. Thus $p_{iat}>0$ should only be introduced in regions where there is knowledge that MMC-Ts were implemented in TMIC settings. 

Taken together, the probability of circumcision of any type, at age $a$ and time $t$ given they were uncircumcised by age $a-1$ and time $t-1$ was 
\begin{equation}
	\begin{split}
	\lambda^{\text{MC}}_{iat} &= \mathbb{P}(\text{MC} \; \text{in} \; (i,a,t) \; | \; \text{Uncircumcised in} \; (i,a-1,t-1)) \\
	              &= \lambda^{\text{TMC}}_{iat} + \lambda^{\text{MMC-T}}_{iat} + \lambda^{\text{MMC-nT}}_{iat} \\ 
	              &= \lambda^{\text{TMIC}}_{ia} + \lambda^{\text{MMC-nT}}_{iat}. 
	\end{split}
	\label{eqn::tot}
\end{equation}
The probability an individual in region $i$ remained uncircumcised at age $a$ and time $t$, given they were uncircumcised by age $a-1$ and time $t-1$, was
\begin{equation}
	\begin{split}
		\lambda^{\text{UC}}_{iat} &= \mathbb{P}(\text{Uncircumcised in} \; \text{in} \; (i,a,t) \; | \; \text{Uncircumcised in} \; (i,a-1, t-1)),\\
		&= 1 - \lambda^{\text{TMIC}}_{iat} - \lambda^{\text{MMC-nT}}_{iat} \\
		&= (1 - \lambda^{\text{TMIC}}_{iat})(1 - \tilde{\lambda}^{\text{MMC-nT}}_{iat})
	\end{split}
	\label{eqn::uncirc}
\end{equation}


\subsubsection{Cumulative incidence of circumcision}


The probability $S_{iat}$ of remaining uncircumcised up to age $a$ and time $t$,  typically referred to as the survivor function, was expressed as
\begin{equation}
	\begin{split}
		S_{iat} &= \mathbb{P}(\text{Uncircumcised in} \; (i,a-1,t-1)) \\
				&= \prod_{(0,(t-a))}^{(a-1,t-1)}\lambda^{\text{UC}}_{iat} \\
				&= \lambda^{\text{UC}}_{i,0,t-a}\cdot \lambda^{\text{UC}}_{i,1,t-a-1}\cdot\ldots \cdot\lambda^{\text{UC}}_{i,a-1,t-1}
	\end{split}
	\label{eqn::survfunc}
\end{equation}
The cumulative incidence function (CIF) defines the marginal probability (or proportion/coverage of) individuals who were circumcised by type $k\in K = \{$ MMC, MMC-nT, MMC-T, TMC, TMIC or MMC$\}$ by age $a$ and time $t$, accounting for the competing risk of other circumcision type. This was calculated as the sum of incidence of circumcision by type $k$ at each age up to $a$:
\begin{equation}
	\begin{split}
		\text{CIF}_{iat}^k &= \mathbb{P}(k \; \text{by} \; (i,a,t)) \\
		&= \sum_{(0,(t-a))}^{(a,t)} I^k_{iat}
	\end{split}
	\label{eqn::cuminc}
\end{equation}
where $I^k_{iat}$ is the incidence of circumcision type $k$ in region $i$, age $a$ and time $t$. For types $k \in \{\textrm{TMIC}, \textrm{TMC}, \textrm{MMC-T}\}$, this was defined by the probability of remaining uncircumcised by any type up to age $a$ at time $t$ times the probability of circumcision by type $k$ in year $t$ at age $a$
\begin{equation}
	\begin{split}
		I_{iat}^k &= \mathbb{P}(k \; \text{in} \; (i,a,t)) \\
		       &= \mathbb{P}(k \; \text{in} \; (i,a,t) | \; \text{Uncircumcised in} \; (i,a-1,t-1))\times \\
		       & \;\;\;\;\;\;\;\;\;\;\;\;\;\;\;\;\;\mathbb{P}(\text{Uncircumcised in} \; (i,a-1,t-1)) \\
	    	   &= \lambda_{iat}^k \cdot S_{iat} 
	\end{split}
	\label{eqn::inc}
\end{equation}
The overall incidence and CIF by any circumcision type can be obtained by summing across all circumcision types. The CIF is also often referred to as the `sub-distribution function', due to the fact that the cumulative probability of a particular event $k$ by time $t$ will remain below one \citep{putter2006tutorial}. 


\subsection{Priors}


In space, we assigned the random effects $\hat{\psi}_i$, $\bar{\psi}_i$, $\psi_i$, $\hat{\gamma}_{ia}$, $\gamma_{ia}$ and $\zeta_{it}$, intrinsic conditional autoregressive (ICAR) priors \citep{besag1995conditional}. An ICAR model encodes spatial dependence between neighbours, defined as areas that share a common boundary, and allows information to be borrowed across regions, which may be useful in areas where data is sparse or non-existent. For a generic parameter $\beta_i$, an ICAR model assumes that the expected value of $\beta_i$ is a weighted average its neighbours, 
\begin{equation*}
	\beta_i \;|\; \beta_{j}, j \sim i, \tau_{\beta} \sim \text{N}\left(\frac{1}{n_i} \sum_{j \sim i} \beta_j, \frac{1}{n_i\tau_{\beta}} \right) \;\;\; i = 1, 2, 3,\ldots, N_I
\end{equation*}
where $j \sim i$ refers to the neighbours of region $i$, $n_i$ is the number of neighbours and $\tau_{\beta}$ is the marginal precision. The joint distribution may be equivalently expressed as $\boldsymbol{\beta} \sim N(\boldsymbol{0}, \tau_{\beta}^{-1}Q^{-1}_{I})$ where $Q_{I}$ is the precision matrix encoding the adjacency structure of the neighbourhoods. The precision matrix $Q_{I}$ is rank deficient and consequently the prior is improper \citep{rue2005gaussian}.

In time, we assigned the random effects $\theta_t$, $\delta_{at}$ and $\zeta_{it}$ an autoregressive process of order 1 (AR1) priors,
\begin{align*} 
  \beta_t \; | \; \rho_{\beta}, \tau_{\beta} &\sim
  \begin{cases}
  	N(0, \tau^{-1}_{\phi}) & a = 0\\
  	N(\rho_{\beta} \cdot \beta_{t-1}, \;\; \tau^{-1}_{\beta}) \;\;\; & t = 1, 2,\ldots, N_A
  \end{cases}
\end{align*}
where $|\rho_{\phi}| < 1$ is an autocorrelation parameters controlling the correlation of the effect of the current age on the previous age and $\tau_{\beta}$ is the precision. The joint distribution may be equivalently expressed as $\boldsymbol{\beta}\; | \; \rho_{\beta}, \tau_{\beta} \sim N(\boldsymbol{0}, \tau_{\beta}^{-1}Q^{-1}_{T}(\rho_{\beta}))$ where $Q_{T}$ is the precision matrix and encodes the first order dependency in time controlled by the autocorrelation parameter $\rho_{\beta}$.

In age, we modelled the random effects $\hat{\phi}_a$, $\bar{\phi}_a$, $\phi_a$, $\hat{\gamma}_{ia}$, $\bar{\gamma}_{ia}$, $\gamma_{ia}$ and $\delta_{at}$ using penalised B-spline functions of the form,
\begin{align*} 
	\beta_a = \sum_{j = 1}^{N_J} \omega_{j}f_{aj}
\end{align*} 
where $f_{aj}$ are B-spline basis functions with knots every five years, $N_J$ is the number of splines used and $\omega_{a}$ are the spline weight parameters to be estimated. The weights $\omega_{a}$ are penalised using an AR1 process (as above) and the joint distribution is given by $\boldsymbol{\omega}\; | \; \rho_{\omega}, \tau_{\omega} \sim N(\boldsymbol{0}, \tau_{\omega}^{-1}Q^{-1}_{A}(\rho_{\omega}))$ where $Q_{A}$ is the precision matrix as defined by the smoothed function. Then defining $\boldsymbol{\beta} = W_{\beta}\cdot \boldsymbol{\omega}$ where $W_{\beta}$ is a design matrix evaluating the basis functions $f_{aj}$ at each age of interest, the joint distribution of $\boldsymbol{\beta}\; | \;\boldsymbol{\omega}, \; W_{\beta}, \; \rho_{\omega}, \tau_{\omega}$ is given by, $\boldsymbol{\beta}\; | \;\boldsymbol{\omega}, \; W_{\beta}, \; \rho_{\omega}, \tau_{\omega} \sim N(\boldsymbol{0}, \tau_{\omega}^{-1}(W_{\beta}\cdot Q^{-1}_{A}(\rho_{\omega})\cdot W_{\beta}^T))$

Moreover, the age-space, age-time, and space-time interaction terms ($\hat{\gamma}_{ia}$, $\bar{\gamma}_{ia}$, $\gamma_{ia}$, $\delta_{at}$ and $\zeta_{it}$) are modelled as Type IV interactions as defined by Knorr-Held and Leonard \citep{knorr2000bayesian},
\begin{align*} 
  \hat{\boldsymbol{\gamma}} \; | \; \rho_{\boldsymbol{\hat{\gamma}}}, \tau_{\boldsymbol{\hat{\gamma}}} &\sim N(\boldsymbol{0}, \;\tau^{-1}_{\hat{\gamma}} Q^{-1}_{\hat{\gamma}}) \\
  \bar{\boldsymbol{\gamma}} \; | \; \rho_{\boldsymbol{\bar{\gamma}}}, \tau_{\boldsymbol{\bar{\gamma}}} &\sim N(\boldsymbol{0}, \;\tau^{-1}_{\bar{\gamma}} Q^{-1}_{\bar{\gamma}}) \\
  \boldsymbol{\gamma} \; | \; \rho_{\gamma}, \tau_{\gamma} &\sim N(\boldsymbol{0}, \;\tau^{-1}_{\gamma} Q^{-1}_{\gamma})\\
  \boldsymbol{\delta} \; | \; \rho_{1\delta}, \rho_{2\delta}, \tau_{\delta} &\sim N(\boldsymbol{0}, \;\tau^{-1}_{\delta} Q^{-1}_{\delta})\\
  \boldsymbol{\zeta} \; | \; \rho_{\zeta}, \tau_{\zeta} &\sim N(\boldsymbol{0}, \;\tau^{-1}_{\zeta} Q^{-1}_{\zeta})
\end{align*}
where the precision matrices constructed using Kronecker products 
\begin{align*} 
	Q_{\hat{\gamma}} &= [W_{\hat{\gamma}}\cdot Q_A(\rho_{\hat{\gamma}}) \cdot W^T_{\hat{\gamma}} ]\otimes Q_I\\
	Q_{\bar{\gamma}} &= [W_{\bar{\gamma}}\cdot Q_A(\rho_{\bar{\gamma}}) \cdot W^T_{\bar{\gamma}} ]\otimes Q_I\\
	Q_{\gamma} &= [W_{\gamma}\cdot Q_A(\rho_{\gamma}) \cdot W^T_{\gamma} ]\otimes Q_I\\
	Q_{\delta} &=[W_{\delta}\cdot Q_A(\rho_{1\delta}) \cdot W^T_{\delta} ] \otimes Q_T(\rho_{2\delta})\\
	Q_{\zeta} &=Q_I \otimes Q_T(\rho_{\zeta})
\end{align*}
Here, $Q_I$, $Q_A$ and $Q_T$ are precision matrices discussed above.

Gaussian priors were assigned to each of the intercepts, $\alpha, \;\hat{\alpha}, \;\bar{\alpha} \sim N(0, 5^2)$. Exponential priors were placed on standard deviations, $\sigma_{(\cdot)} = \sqrt{\tau^{-1}_{(\cdot)}} \sim \text{Exp}(1)$. Gaussian priors were set for all correlation parameters on the logit scale, such that 
\begin{align*} 
  \hat{\rho}_{(\cdot)} = \frac{2}{1 + \exp(-\rho_{(\cdot)})} - 1 \sim N(3, 1^2)
\end{align*}
corresponding to 95\% CI prior weight that the autocorrelation parameters are between 0.48 and 0.99 on the real scale. The proportion of TMICs performed as MMC-Ts, $p_{iat}$, were assigned independent Gaussian distributions on the logit scale, 
\begin{equation*} 
  	\text{logit}(p_{iat}) \sim N(\mu_{iat}, \sigma_{iat}^2) 
\end{equation*}
if there is evidence or knowledge of the VMMC programmes to suggest MMC-Ts are taking place with some mean, $\mu_{iat}$, and standard deviation, $\sigma_{iat}$. Otherwise, the proportions of TMICs performed as MMC-Ts are fixed at zero ($p_{iat}$=0).


\subsection{Likelihood}


The full likelihood function consists of the product of the likelihoods for two independent data sources: (1) household survey data on individual male's age at circumcision and circumcision type, and (2) VMMC programme data about the number of circumcisions conducted for HIV prevention in each region.


\subsubsection{Survey data}


Survey data consisted of individual observations of self-reported age at circumcision, date of birth, and circumcision type (medical or traditional) for male survey respondents. Circumcision status observations were right censored if the respondent reported not being circumcised at the time of the survey or left censored if the individual reported being circumcised at the time of survey but did not report their age at circumcision.

Each survey  $s \in S = \{1, 2, \ldots, N_S\}$ consisted of a sample of individuals $j \in J_s = \{0, 1, 2, \ldots, N_{J_s}\}$ residing in region $i \in I = \{1, 2, \ldots, N_I\}$. Surveyed individuals were followed from their year of birth until their year of circumcision or their year of censoring. For circumcised individuals, the year of circumcision was calculated as the year of birth plus the age at circumcision. It was assumed that no circumcisions occurred after age 59, so for uncircumcised individuals, censoring occurred either in the survey year or when they were 59. Individuals who self-reported they were circumcised but have a missing age at circumcision were included in the analysis through left censoring. 

The partial likelihood contribution for each individual is this defined by the following groups:
\begin{itemize}
	\item \textbf{TMIC observed}: Individuals in region $i$ reporting they were circumcised through TMI ceremonies aged $a$ at time $t$. The likelihood of this outcome was the probability an individual in region $i$ reported a TMIC at age $a$ and time $t$, $I^{\text{TMIC}}_{iat} = \lambda^{\text{TMIC}}_{iat}\cdot S_{iat}$. 
	\item \textbf{MMC-nT observed}: Individuals in region $i$ reporting they were medically circumcised aged $a$ at time $t$. The likelihood of this outcome was the incidence of MMC-nT in region $i$, at age $a$ and time $t$, $I^{\text{MMC-nT}}_{iat} = \lambda^{\text{MMC-nT}}_{iat}\cdot S_{iat}$. 
	\item \textbf{Right censored}: Individuals in region $i$ who were uncircumcised by age $a$ at time $t$.  The likelihood of this outcome was the probability of remaining uncircumcised in region $i$, at age $a$ and time $t$,$S_{iat}$
	\item \textbf{Left censored}: Individuals in region $i$ reporting a circumcision at an unknown age and time between birth at time $t-a$ and age $a$ at time $t$. The likelihood of this outomce was the probability of being circumcised in region $i$ before age $a$ at time $t$, $(1-S_{iat})$ 
\end{itemize}

Taken together, the partial likelihood for the survey data may be expressed as 
\begin{eqnarray*}
	L(\boldsymbol{\Theta}) &= \displaystyle\prod_{(i,a,t)} \prod_{s=1}^{N_s} \prod_{j=1}^{J_s}  \underbrace{\left(\lambda^{\text{TMIC}}_{iat}\cdot S_{iat}\right)^{\mathbbm{1}^{\text{TMIC}}_{sjiat}}}_{\text{TMIC observed}} \cdot  \underbrace{ \left(\lambda^{\text{MMC-nT}}_{iat} \cdot S_{iat}\right)^{\mathbbm{1}^\text{MMC-nT}_{sjiat}}}_{\text{MMC-nT observed}} \cdot
	  \underbrace{\left(S_{iat}\right)^{\mathbbm{1}^{RC}_{sjiat}}}_{\text{Right censored}} \cdot \underbrace{\left(1-S_{iat}\right)^{\mathbbm{1}^{LC}_{sjiat}}}_{\text{Left censored}} \\
	   &= \displaystyle\prod_{(i,a,t)} \left(S_{iat}\lambda^{\text{TMIC}}_{iat}\right)^{\text{N}^{\text{TMIC}}_{iat}} \cdot \left(S_{iat}\lambda^{\text{MMC-nT}}_{iat}\right)^{\text{N}^\text{MMC-nT}_{iat}} \cdot
       \left(S_{iat}\right)^{\text{N}^{RC}_{iat}} \cdot \left(1-S_{iat}\right)^{\text{N}^{LC}_{iat}}
\end{eqnarray*}
where $\mathbbm{1}^{l}_{sjiat}$ is an indicator variable indicating whether individual $j$ in survey $s$ and region $i$ reported outcome $l$ at age $a$ and time $t$. Here, $\text{N}^{l}_{sjiat}$ denoting the total number of individuals reporting outcome $l$ in region $i$ at age $a$ and time $t$
\begin{equation*}
	\text{N}^{l}_{iat} = \sum_{s=1}^{N_s}\sum_{j=1}^{J_s} \mathbbm{1}^k_{sjiat}
\end{equation*}
where $l$ is TMIC, MMC-nT, RC or LC.

Survey data are often collected through a complex two-stage cluster sampling design with unequal sampling probabilities. As a result, performing model inference using a standard likelihood may lead to biased estimates. To account for these survey designs, the probabilities of circumcision were estimated using a weighted pseudo-likelihood in which we replaced the observed counts $\text{N}^l_{iat}$ with weighted counts $\tilde{\text{N}}^l_{iat}$, calculated using survey weights. Each individual $j$ in survey $s$ residing in region $i$ had sampling weight $\omega_{sji}$ which were normalised using the Kish effective sample size, 
\begin{equation*}
	\tilde\omega_{sji} = \frac{\omega_{sji}}{\bar{\omega}_{si}}\cdot \frac{M_s}{M^{\text{eff}}_s} 
\end{equation*}
Here, $\bar{\omega}_{si}$ is the arithmetic mean of all survey weights from the individuals sampled in survey $s$ and region $i$ and $M_s$ is the total sample size in survey $s$ and $M^{\text{eff}}_s$ is the Kish effective sample size which accounts for heterogeneity in sampling weights and is calculated using 
\begin{equation*}
	M^{\text{eff}}_s = \frac{(\sum_j \omega_{sji})^2}{\sum_j \omega_{sji}^{2}}.
\end{equation*}
Using the normalised sampling weights, $\tilde\omega_{sji}$, we calculate the weighted counts, $\tilde{\text{N}}^l_{iat}$, for each region, age and time stratum using 
\begin{equation*}
	\tilde{\text{N}}^{l}_{iat} = \sum_{s=1}^{N_S}\sum_{j=1}^{N_{J_s}}\tilde\omega_{sji} \mathbbm{1}^{\text{l}}_{sjiat}.
\end{equation*}
where $l$ is either TMIC, MMC-nT, RC or LC. 


\subsubsection{VMMC programme data}


Data on the number of VMMCs conducted by public health programmes consisted of the total number of MMCs (both MMC-nT and MMC-T) conducted in region $i$ at time $t$ and age group $G=[a_1,a_2]$, $Y_{iGt}$. We modelled number of MMCs, $Y^{\text{MMC-nT}}_{iat}$, and MMC-Ts, $Y^{\text{MMC-T}}_{iat}$ in region $i$, at age $a$ and time $t$ using a Poisson likelihood with means $\mu^{\text{MMC}}_{iat}$ and $\mu^{\text{MMC-T}}_{iat}$ respectively,
\begin{equation*}
    Y^{k}_{iat} \;|\; \mu^{k}_{iat}\sim \text{Poisson}(\mu^{k}_{iat}) \\
	\label{eqn::VMMCPoisson}
\end{equation*}
where $k$ is MMC-T or MMC-nT. The mean $\mu^{k}_{iat}$ is determined by multiplying the male population size in region $i$, at age $a$ and time $t$, $P_{iat}$, the probability of remaining uncircumcised in region $i$, up to age $a$ and time $t$, $S_{iat}$, and the probability of being having a MMC-T or MMC-nT at age $a$ and time $t$ given they were uncircumcised by age $a-1$ and time $t-1$, $\lambda^{\text{MMC-T}}_{iat}$ or $\lambda^{\text{MMC-T}}_{iat}$
\begin{equation*}
    \mu^k_{iat} = P_{iat}\cdot S_{iat} \cdot \lambda^{k}_{iat}.
\end{equation*}
where $k$ is MMC-T or MMC-nT. 
\noindent It then follows that the total number of MMCs conducted in region $i$ at time $t$ and age group $G=[a_1,a_2]$, $Y_{iGt}$ are modelled using 
\begin{equation*}
	\begin{split}
		Y_{iGt} \;|\; \mu_{iat} &\sim \text{Poisson}(\mu_{iGt}) \\
		\mu_{iGt} &= \sum_{a = a_1}^{a_2} \mu^{\text{MMC-nT}}_{iat} + \mu^{\text{MMC-T}}_{iat}.
	\end{split}
\end{equation*}


\subsection{Inference}


Models were implemented and fitted in R \citep{rcore} using Template Model Builder (TMB) \citep{kristensen2016tmb}. TMB is a software package that enables users to flexibly fit latent process/variable models to data. It uses automatic differentiation and Laplace approximations to estimate posterior distributions for model parameters. Models were optimised using the quasi-newton L-BFGS-B optimisation method \citep{byrd1995limited}.

Predictions of the district-age-time probabilities of circumcision and other quantities of interest were estimated using Monte Carlo sampling, drawn from the joint posterior distribution, conditional on the optimised hyper-parameters \citep{eaton2021naomi}. Marginal predictive distributions of any quantity of interest, for example, circumcision coverage aggregated to any geographical unit and/or age groups can then be made by summarising the joint samples generated.


\section{Application to male circumcision coverage in South Africa}


In this section, we applied the methods described in Section \ref{sec::methods} to estimate annual probabilities of becoming circumcised and the corresponding circumcision coverage in South Africa between 2008 and 2019, by type, by single-year age group, and at a district-level (admin-2). 


\subsection{Data and outputs}\label{sec::data}


We included nationally-representative household survey data on self-reported circumcision status from five surveys conducted in South Africa between 2002 and 2017: the South African National HIV Prevalence, HIV Incidence, Behaviour and Communication Survey (SABSSM) from 2002, 2008, 2012 and 2017 \citep{hsrc2017, hsrc2012, hsrc2008, hsrc2002} and the South Africa Demographic and Health Survey (DHS) 2016 \citep{dhs2016}. Information related to age, residence, self-reported circumcision status, age at circumcision, who performed the circumcision and where the circumcision took place were extracted for 51,261 male respondents across all surveys. The specific questions about circumcision status in each survey are in Supplementary Material Table B.1. Circumcisions were classified as either MMC-nT or TMIC by using responses to both `Who performed the circumcision?' and `Where did the circumcision take place?' using the criteria described in Supplementary Material Table B.2.

We obtained data on the number of VMMCs performed annually among men aged 10 years and older from South Africa NDoH District Health Information System (DHIS) for South African government fiscal years (April to March) from 2013 through 2019. These were supplemented for the years 2009 through 2012 by district-level data on number of circumcisions conducted recorded in the DMPPT2 model applications \citep{kripke2016age, kripke2016cost}. For 2018 through 2019, the number of VMMCs conducted in districts in the Eastern Cape province were sourced from data reported to PEPFAR via the Monitoring, Evaluation, and Reporting (MER) Indicator framework instead of NDoH DHIS due to discrepancies in the reporting of MMC-Ts between PEPFAR MER and NDoH DHIS. The total number of VMMCs performed in each district by data source is shown in Supplementary Material Figure B.4. Further details on each data source can be found in the Supplementary Material. 

South Africa is composed of 52 districts (admin-2) situated in nine provinces (admin-1). Annual district-level estimates for the male population size by five-year age group from 2008 through 2019 were sourced from Statistics South Africa Mid-Year Population Estimates 2020 \citep{StatsSAPop}. District population estimates were scaled to align to provincial population estimates by five-year age group from the Thembisa version 4.4 HIV and demographic model, and distributed to single-year of age according to the proportion of population within each single-year age group from Thembisa v4.4 \citep{johnson2021thembisa}. 

The model produced the resulting annual probabilities of becoming circumcised and the corresponding circumcision coverage in South Africa between 2008 and 2019, by circumcision type, by single-year age group, and at a district-level. Results were aggregated: (1) from district level to coarser administrative boundaries (province, national), (2) from single-year age groups to five-year age group (0-4, 5-9, etc.) as well as coarser priority age groups (15-49, 15-29 etc.) and (3) from individual types of circumcision (MMC-nT, MMC-T and TMC) to produce combinations including MMC, TMIC and MC. For each output, the posterior mean, median, standard deviation, and quantile-based 95\% credible intervals (CI) are computed. A full list of model inputs and output can be seen in Supplementary Material Table A.1. Table 1 shows the MC, MMC and TMC coverage, along with the changes in MC coverage between 2008 and 2019 and the breakdown of circumcised and uncircumcised men for men aged 15-49 years by district, province and nationally. A full list of model inputs and output are summarised in Supplementary Material Table A.1. 


\subsection{Assumptions about circumcisions in TMIC}


Direct data about MMCs conducted in traditional male initiation contexts (MMC-T) were not available. Consequently, strong prior assumptions are required about the proportion of circumcisions conducted in TMIC which were MMC-T, informed by expert knowledge of the national circumcision programme. Within the model, it was assumed that all TMICs are TMC prior to 2013, after which a proportion of TMIC circumcisions conducted were assumed to be MMC-Ts as follows:
\begin{itemize}
	\item In Nkangala District in Mpumalanga Province, we assumed that around 90\% of the TMICs were MMC-Ts each year from 2013 to 2019.
	\item In Waterberg, Capricorn, Vhembe, Sekhukune, Mopani, Ehlanzeni and Gert Sibande districts (Limpopo and Mpumalanga Provinces), we assumed that an increasing proportion of TMICs were done in a medical context between 2015-2019. Around 20\% (in 2015), 40\% (in 2016), 60\% (in 2017), 80\% (in 2018), 90\% (in 2019) TMICs are assumed to be MMC-Ts each year.
	\item For districts in Eastern Cape province and the City of Cape Town district, we assumed that 99\% of men receiving TMICs were MMC-Ts in 2018 and 2019. For the City of Cape Town, the majority of men receiving TMIC were Xhosa who returned to traditional male initiation ceremonies in Eastern Cape province. 
	\item In all other districts, it is assumed that there were no TMICs done in a medical context.
\end{itemize}

Many young men who migrate for work, for example Xhosa men from the Eastern Cape province to Cape Town and other parts of the Western Cape, return to their family home for traditional male initiation. Such men appear in household surveys in their district of residence, and the model estimates the probabilities of circumcision among men in their district of residence, not where the circumcision occurred. However, MMC-T provided among these men during TMIC would be recorded in the district where the ceremony occurred. To account for this, the number of VMMCs reported in Eastern Cape province in 2018 and 2019 were re-allocated to all districts in South Africa proportionally to the distribution of men reporting isiXhosa as their primary language in the South Africa 2011 census. The number of re-allocated circumcisions were small in most districts, with most reallocated to larger metropolitan areas such as Cape Town and Johannesburg (Supplementary Material Figure B.4). See Supplementary Material for more details. 


\subsection{Results}

\begin{figure}[H]
	\centering
	\includegraphics[width = 0.85\linewidth]{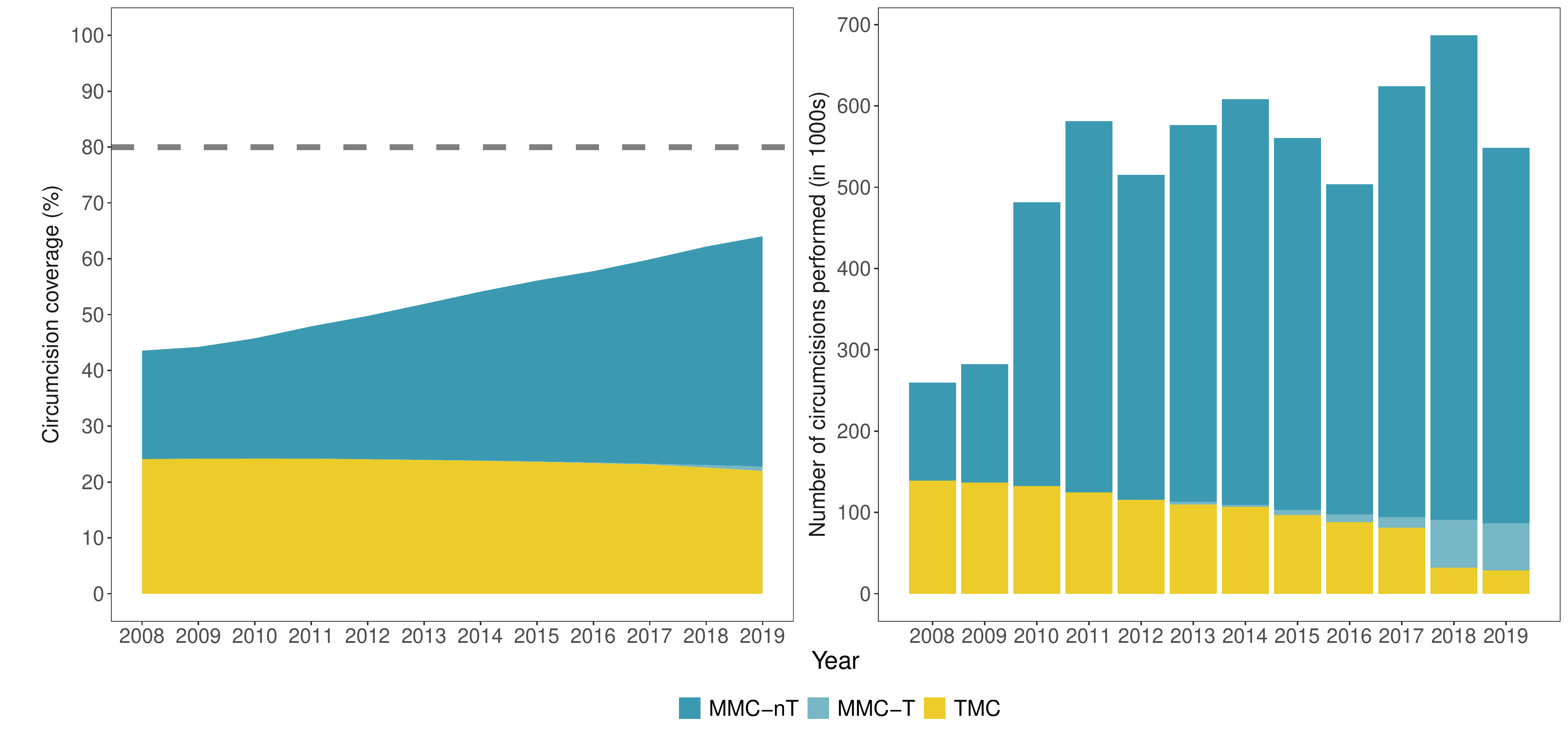}
	\caption{(Left) Estimated MC coverage nationally among men aged 15-49 years between 2008 and 2020 disaggregated by circumcision type. Lines represent the posterior mean, with the dashed line denote the target circumcision coverage of 80\%. (Right) Estimated number of circumcisions performed annually between 2010 and 2020 disaggregated by type. Bars represent the posterior mean.}
	\label{fig::1549prev}
\end{figure}

Nationally in South Africa, between 350,000 and 650,000 VMMCs were conducted annually between 2010 and 2019 (Figure \ref{fig::1549prev}). As a result, MC coverage in men aged 15-49 years was 64.0\% (95\% CI: 63.2\% to 64.9\%) in 2019, an increase of 20.5\% (95\% CI: 20.0\% to 21.0\%) since 2008. When stratifying by circumcision type, traditional circumcision was the most prevalent among circumcised men aged 15-49 years in 2008 with a TMC coverage of 24.1\% (95\% CI: 23.4\% to 24.8\%) compared with an MMC coverage of 19.4\% (95\% CI: 18.9\% to 20.0\%). This reversed by 2019 with and MMC coverage of 42.0\% (95\% CI: 41.3\% to 43.0\%) among men aged 15-49 years compared with a decrease in TMC coverage to 22.0\% (95\% CI: 21.3\% to 22.7\%).

\begin{figure}[H]
	\centering
	\includegraphics[width = 0.85\linewidth]{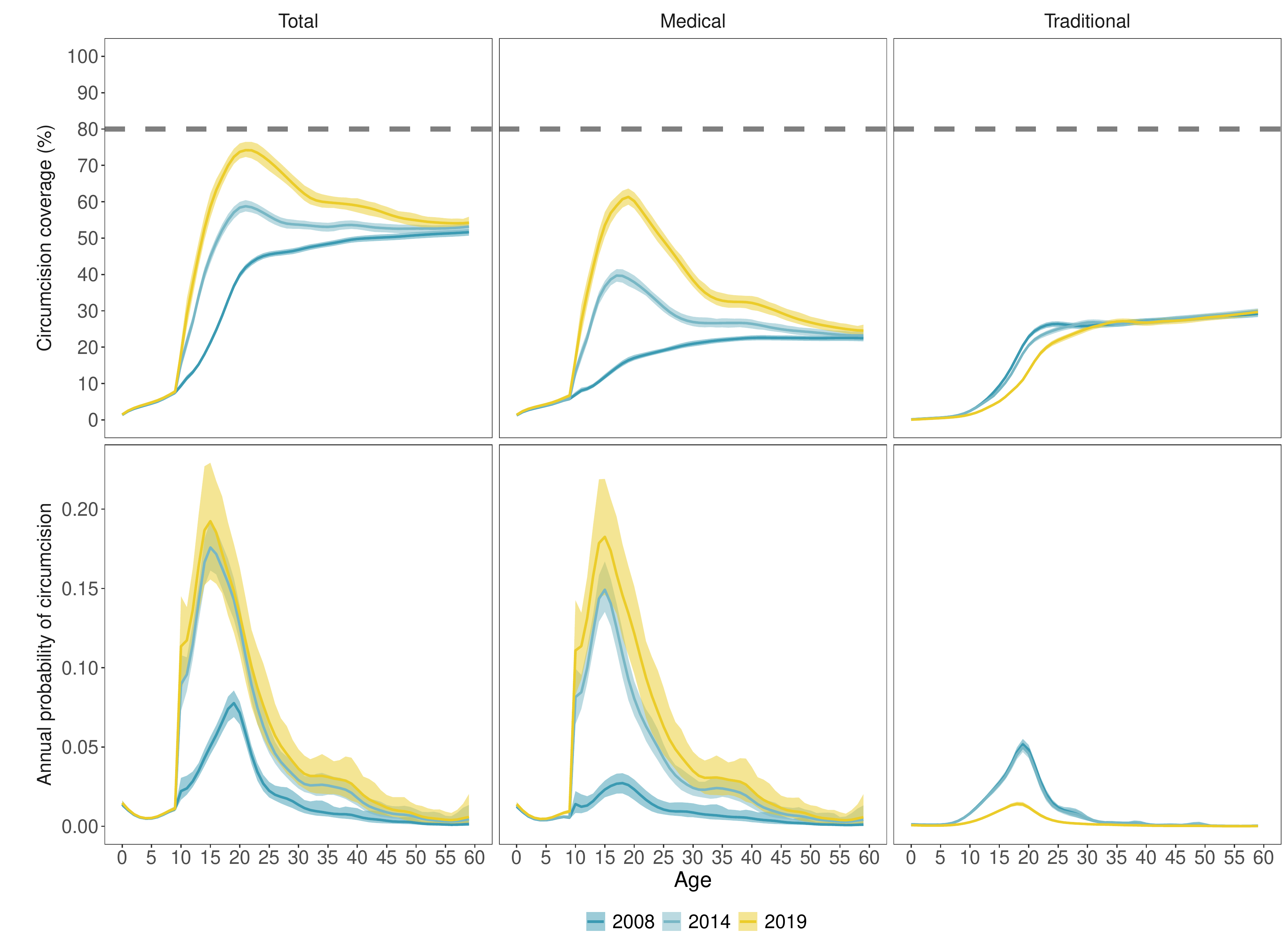}
	\caption{(Top) Estimated Total (MC), Medical (MMC) and Traditional (TMC)  coverage nationally by age in 2008, 2014 and 2020. (Bottom) Estimated probability of Total (MC), Medical (MMC) and Traditional (TMC) circumcision nationally by age in 2008, 2014 and 2020. Lines denote the posterior mean with shaded regions denoting the 95\% CI. Dashed line denotes the target circumcision coverage of 80\%.}
	\label{fig::singleageprev}
\end{figure}

The largest increase in MC coverage between 2008 and 2019 was among men aged 10-25 years (Figure \ref{fig::singleageprev}), ages specifically targeted by the South African VMMC programme. MC coverage peaked at 74.2\% (95\% CI: 72.3\% to 76.5\%) in men aged 21 years in 2019, an increase of 32.2\% (95\% CI: 30.3\% to 34.4\%) since 2008. The large increases in circumcision coverage in men aged 10-25 years were due to the increases in medical circumcision. The highest MMC coverage in 2019 was 61.3\% (95\% CI: 59.2\% to 63.6\%) in men aged 19 years, an increase of 44.9\% (95\% CI: 42.6\% to 47.3\%) between 2008 and 2019. Conversely, TMC coverage among men aged 10-25 years has decreased with the corresponding TMC coverage in men aged 19 of 11.0\% (95\% CI: 10.6\% to 11.5\%) in 2019, a decrease of 9.4\% (95\% CI: 9.0\% to 9.9\%) since 2008.



\begin{figure}[H]
	\centering
	\includegraphics[width = 0.75\linewidth]{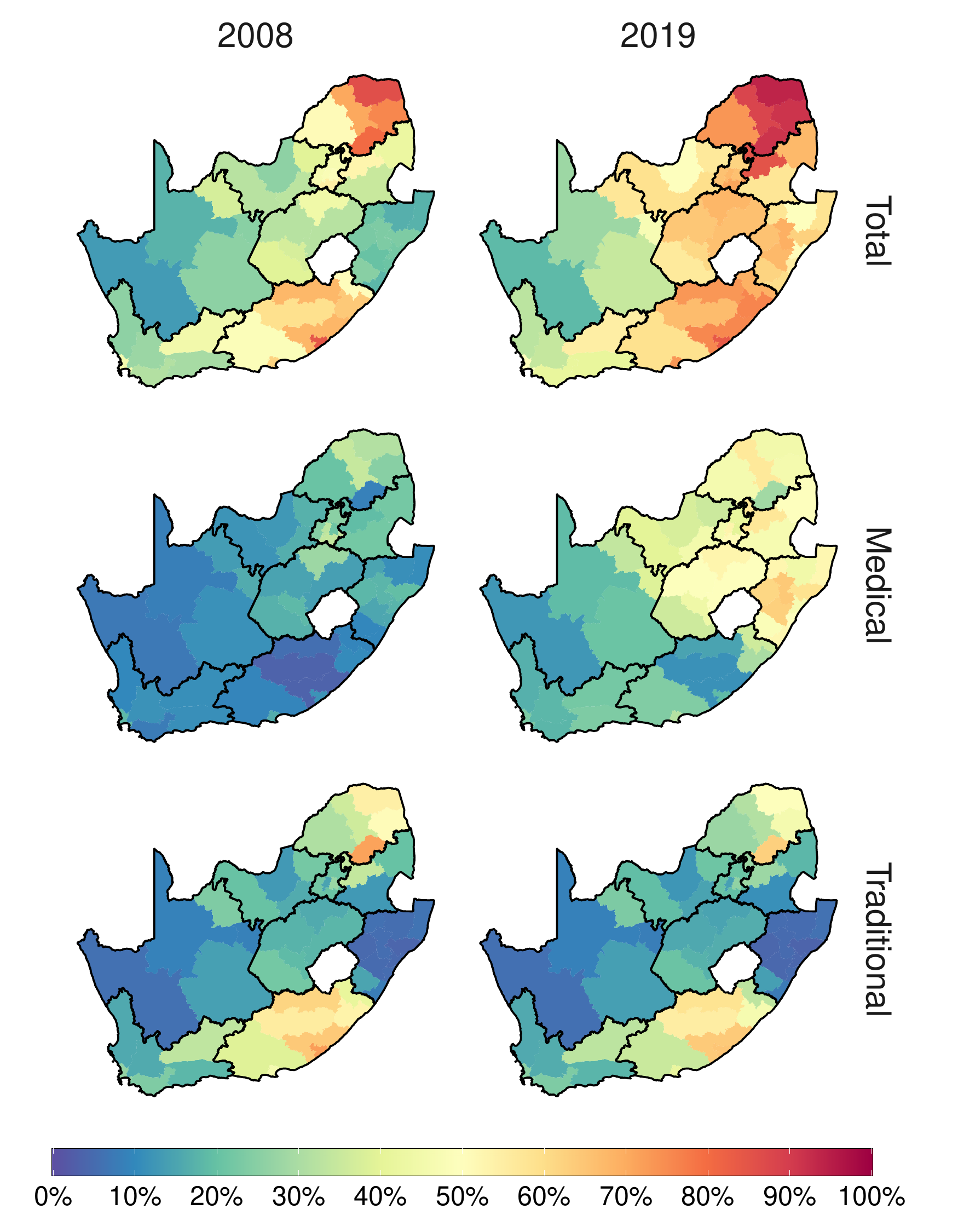}
	\caption{Estimated Total (MC), Medical (MMC) and Traditional (TMC) circumcision coverage for men aged 15-49 in each district in 2008, 2014 and 2020. Colours denote the posterior mean.}
	\label{fig::district1549prevmap}
\end{figure}

\begin{figure}[H]
	\centering
	\includegraphics[width = \linewidth]{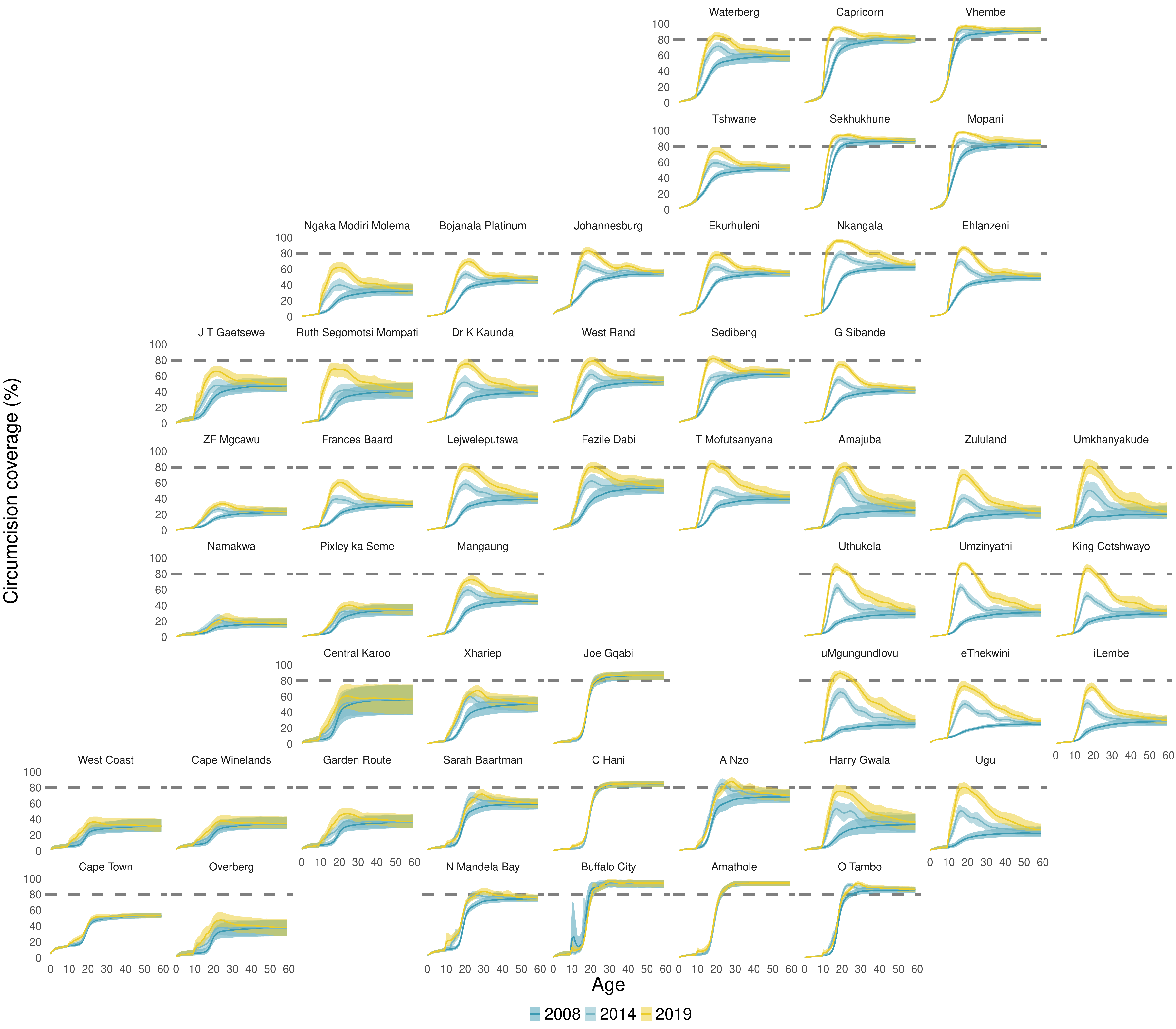}
	\caption{Estimated MC coverage by age and district in 2008, 2014 and 2020. Lines denote the posterior mean with shaded regions denoting the 95\% CI. Dashed lines denote the target circumcision coverage of 80\%. Plots are organised by geography.}
	\label{fig::districtsingleageprev}
\end{figure}

Circumcision coverage varied considerably across South Africa (Figure \ref{fig::district1549prevmap}). In 2008, prior to the scale-up of VMMC, MC coverage was highest in Vhembe, Buffalo City, and Sekhukune districts among men aged 15-49 years with 86.3\% (95\% CI: 81.4\% to 89.8\%), 84.6\% (95\% CI: 77.6\% to 90.7\%) and 80.6\% (95\% CI: 76.8\% to 83.6\%) coverage, respectively. Namakwa, Umkhanyakude, and Zululand districts had the lowest MC coverage, with 13.4\% (95\% CI: 9.4\% to 18.9\%), 16.9\% (95\% CI: 11.7\% to 21.7\%), and 17.4\% (95\% CI: 12.1\% to 24.1\%). This large range in the total circumcision coverage was due to considerably different traditional circumcision practices across South Africa. Districts in Eastern Cape, Limpopo, and Mpumalanga provinces had high levels of traditional circumcision reaching to over 90\% coverage among men by age 30 years in Buffalo City (Figure \ref{fig::districtsingleageprev}). These districts have large populations of Ndebele, Xhosa, Pedi, Venda, and Tsonga people, who all typically perform TMC as a rite of passage. The lowest levels of traditional circumcision were in KwaZulu-Natal, where less than 10\% of men were traditionally circumcised. Medical circumcision coverage was relatively low among men aged 15-49 years in most districts in 2008 ranging between 3.5\% (95\% CI: 1.6\% to 5.5\%) in Chris Hani District and 34.7\% (95\% CI: 29.7\% to 41.2\%) in Capricorn District.

\begin{figure}[H]
	\centering
	\includegraphics[width = \linewidth]{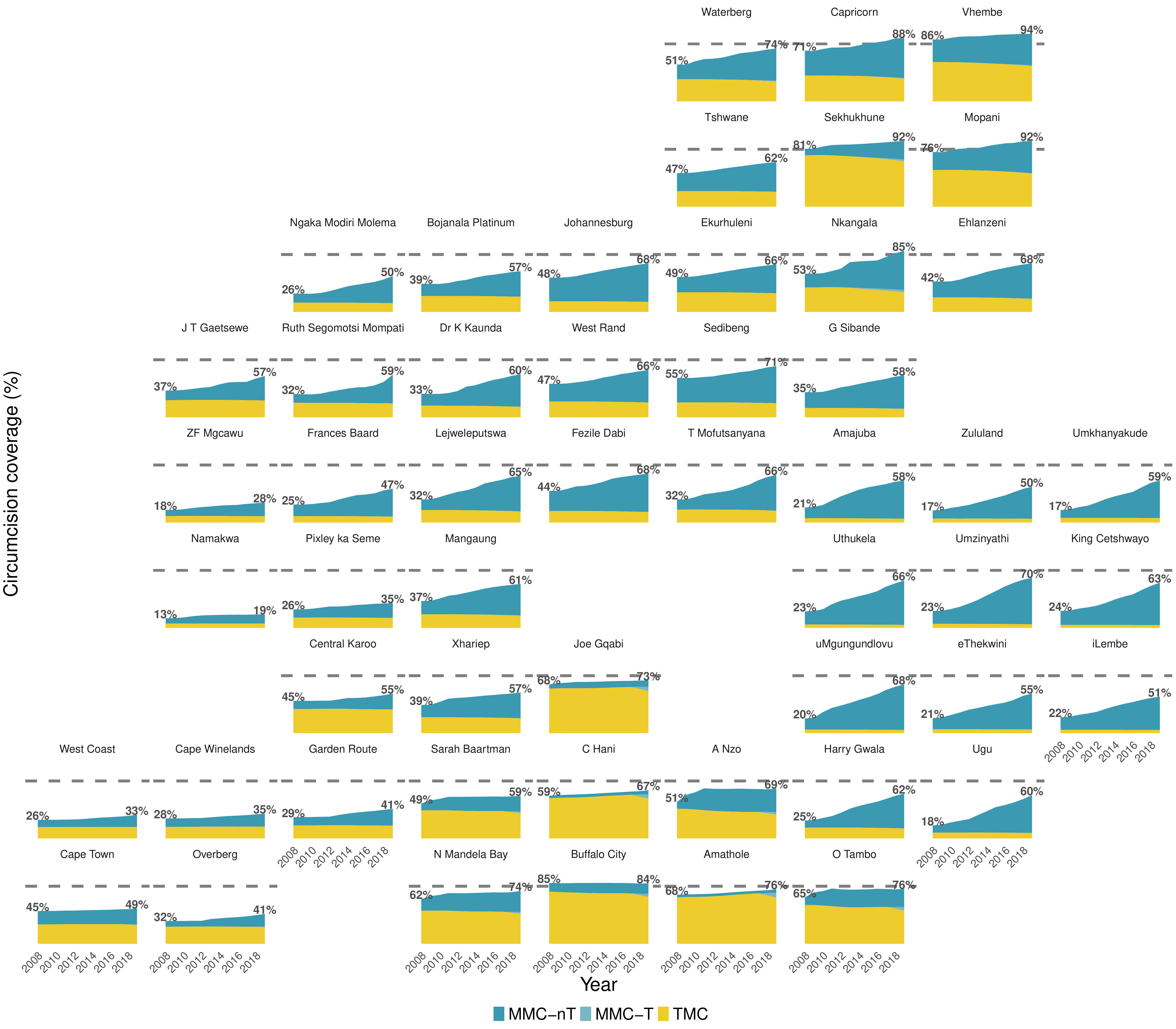}
	\caption{Estimated MC coverage in men aged 15-49 between 2008 and 2020 by district disaggregated by circumcision type. Lines denote the posterior mean. Dashed lines denote the target circumcision coverage of 80\%. Plots are organised by geography.}
	\label{fig::district1549prev}
\end{figure}

Circumcision coverage increased in all districts in South Africa since the implementation of VMMC national campaigns in 2010 (Figure \ref{fig::district1549prev}). Increases have not been uniform between districts as VMMCs have typically been prioritised in areas where HIV incidence and prevalence are high and where there is higher acceptance of VMMC as a HIV prevention \citep{WHOVoluntary2}. Over half of the VMMCs performed were conducted in the three provinces with the highest HIV prevalence among 15-49 year-olds: KwaZulu-Natal (26.8\%), Mpumalanga (25.1\%), and Free State (22.1\%) \citep{hivdata}. Consequently, the districts with the largest changes in MC coverage between 2008 and 2019 were also located in these provinces, with increases of of 47.2\% (95\% CI: 44.9\% to 49.6\%), 46.0\% (95\% CI: 43.4\% to 48.1\%) and 42.9\% (95\% CI: 40.4\% to 45.2\%) in uMgungundlovu, Umzinyathi, and Uthukela districts, respectively. In 2019, Vhembe, Mopani, and Sekhukune districts had the highest MC coverage among men aged 15-49 years with 94.2\% (95\% CI: 90.1\% to 96.7\%), 91.6\% (95\% CI: 88.0\% to 94.6\%) and 91.6\% (95\% CI: 88.9\% to 93.9\%) coverage. Namakwa, ZF Mgcawu, and West Coast district had the lowest MC coverage in 2019 at 18.7\% (95\% CI: 15.2\% to 23.3\%), 28.0\% (95\% CI: 25.3\% to 31.0\%) and 32.9\% (95\% CI: 27.6\% to 39.6\%) coverage. Medical circumcision coverage among men aged 15-49 years ranged considerably in 2019 from 11.6\% (95\% CI: 10.1\% to 13.0\%) in Chris Hani to 64.5\% (95\% CI: 62.2\% to 66.9\%) in Umzinyathi. TMC coverage decreased between 2008 and 2019, with the largest decreases in Sekhukhune, Alfred Nzo and Oliver Tambo of 8.6\% (95\% CI: 7.6\% to 9.6\%), 8.0\% (95\% CI: 6.2\% to 10.0\%) and 7.6\% (95\% CI: 6.4\% to 8.7\%), respectively.



\begin{figure}[H]
	\centering
	\includegraphics[width = 0.65\linewidth]{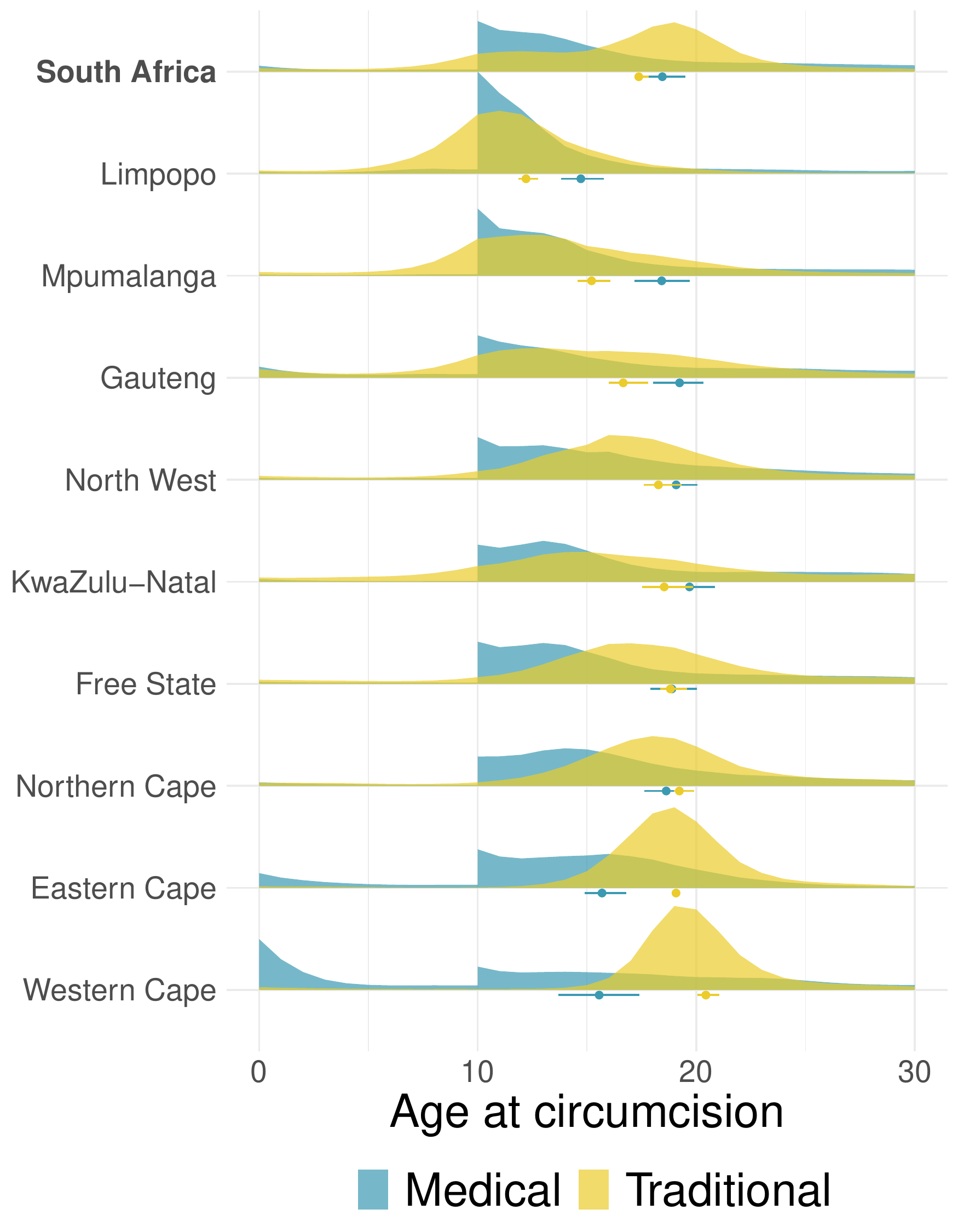}
	\caption{Distribution of the estimated number of medical (MMC-nT) and traditional (TMIC) circumcisions in 2019 by province and for South Africa. Dots denote the average age of circumcision, with the bars denoting the 95\% CI.}
	\label{fig::avgageofcirc}
\end{figure}

The distribution of age at circumcision varied by circumcision type and geography (Figure \ref{fig::avgageofcirc}). Nationally, for circumcisions conducted in 2018, the average age of medical circumcision was 18.4 (95\% CI: 17.4 to 19.5) and for circumcisions in traditional settings was 17.4 (95\% CI: 17.2 to 17.8). The average age of traditional circumcision was the lowest in Limpopo province for both traditional (12.2 years; 95\% CI: 11.9 to 12.8) and medical circumcision (14.7 years; 95\% CI: 13.8 to 15.8). Mpumalanga province also had a lower average of traditional circumcision than the national average at 15.2 years (95\% CI: 14.6 to 16.1) years for traditional circumcision. In other provinces, traditional circumcision typically occurs around age 18. Western Cape also had a substantially lower average age of medical circumcision (15.6 years; 95\% CI: 13.7 to 17.4) due to the considerable number of circumcisions at birth, which are not related to VMMC programme circumcisions. 



By 2019, only 6 districts (Vhembe, Mopani, Sekhukhune, Capricorn, Nkangala and Buffalo City) out of 52 were estimated to have achieved the target of 80\% MC coverage in men aged 15-49 years (Table \ref{tab:results}) and no districts achieved the new ambitious targets of 90\% MC coverage in adolescent boys and young men aged 10-29 years. However, 12 districts achieved 80\% coverage in the 15-24 years age group (Mopani, Vhembe, Nkangala, Capricorn, Sekhukhune, Umzinyathi, uMgungundlovu, Uthukela, Ehlanzeni, King Cetshwayo, Waterberg, T Mofutsanyana, Johannesburg and Sedibeng), therefore continuing to provide MMC services on this scale will ensure more districts will achieve these ambitious targets in the future. Between 2017 and 2019, over 1.7 million VMMCs were delivered. In order to meet the South African Government's 2.5 million VMMCs target, a further 800,000 VMMCs must be delivered by the end of 2022. In 2019, there were estimated to be 5.43 million (95\% CI: 4.83 to 5.67 million) uncircumcised men aged 15-49 years, with the largest number located in metropolitan areas of Cape Town (623,000; 95\% CI: 562,000 to 659,000), Johannesburg (554,000; 95\% CI: 451,000 to 614,000), and eThekwini (480,000; 95\% CI: 360,000 to 528,000).



\begin{figure}[H]
	\centering
	\includegraphics[width = 0.85\linewidth]{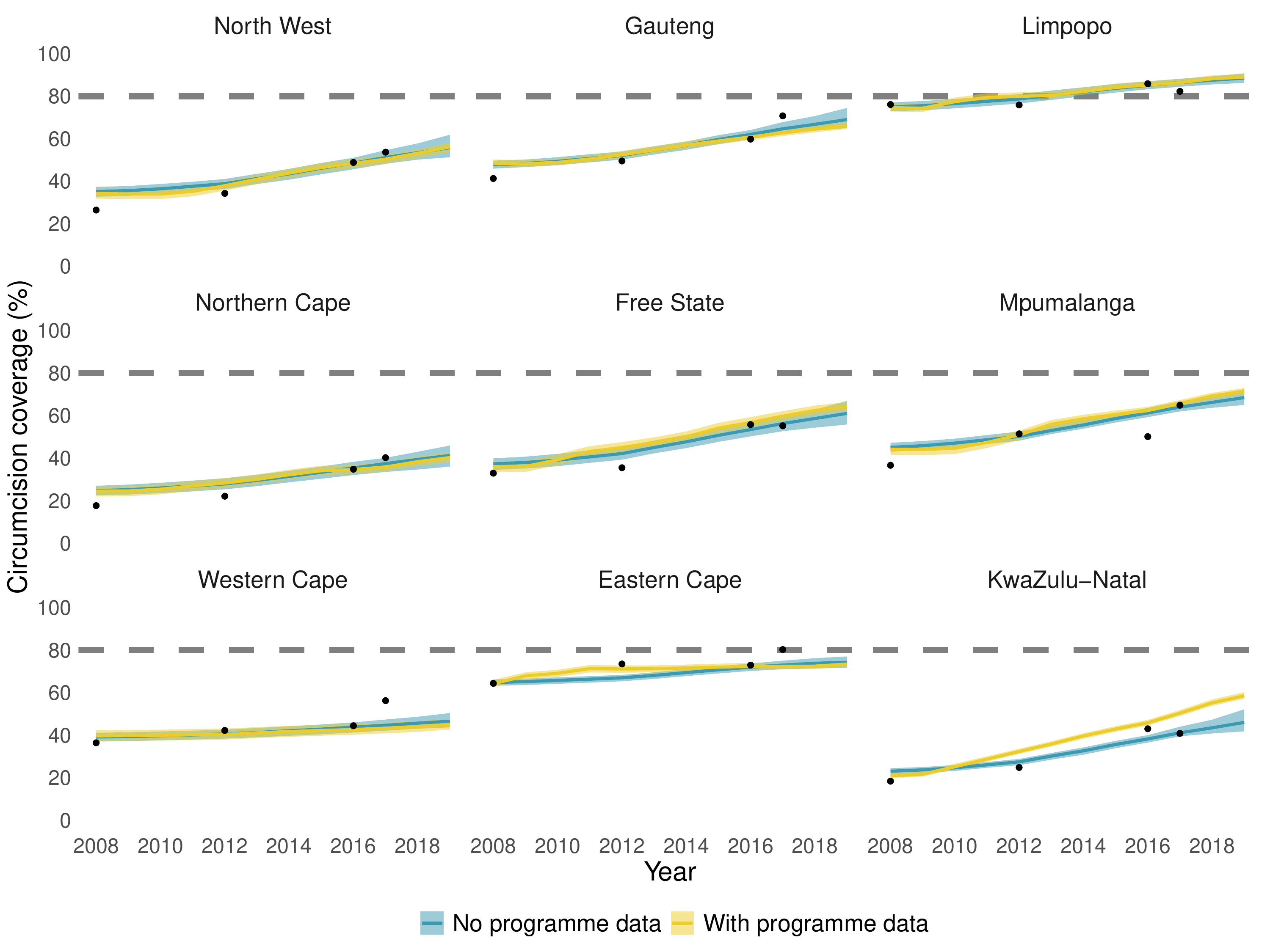}
	\caption{Estimated MC coverage in men aged 15-49 between 2008 and 2019 by province for models with and without programme data. Lines denote the model estimated mean prevalence, with the shaded regions denoting the 95\% CI. Black dots denote the direct survey estimates for MC coverage from the 2008, 2012 and 2017 SABSSM surveys and the 2016 DHS survey.}
	\label{fig::comparison}
\end{figure}

The model presented here utilises both survey and programme data to produce timely estimates of MC coverage, particularly where survey data is limited or no longer available. Estimates of MC coverage were similar in most provinces when produced from models with and without the programme data, except in KwaZulu-Natal where MC coverage was substantially higher by 2019 when including programme data (58.6\%; 95\% CI: 57.1\% to 60.3\%) than excluding programme data (46.0\%; 95\% CI: 41.8\% to 52.2\%) (Figure \ref{fig::comparison}). Overall, national MC coverage was 64.0\% (95\% CI: 63.2\% to 64.9\% in 2019  when including programme data, 1.6\% higher than without program data included in the model (62.4\% (95\% CI: 59.3\% to 66.5\%), indicating a high level of agreement in the estimated number of MMCs conducted from the survey data with the number of MMCs conducted by VMMC program providers in South Africa. 


\section{Discussion}


The ability to produce comprehensive, accurate and timely estimates of MC coverage by age and type of circumcision (traditional or medical) over time at sub-national levels are essential for planning and delivering VMMCs to meet programmatic targets, and evaluating the impact of VMMC campaigns on HIV incidence. In this paper, we developed a model to produce region-age-time-type specific probabilities and the corresponding coverage of MC along with associated measures of uncertainty. The model extends a competing risks time-to-event model in order to integrate both survey data and VMMC programme data, building on previous approaches from \citet{kripke2016cost} and \citet{cork2020mapping}, which did not model both data sources formally. 

The methods described were applied to produce estimates of medical and traditional circumcision coverage in single-year age groups between 2008 and 2019 for each district (admin-2) in South Africa. The estimates highlighted considerable heterogeneity in MC coverage and the changes between 2008 and 2019. Circumcision coverage increased in all districts in South Africa during that period that time, however the largest increases in MC coverage occurred in KwaZulu-Natal province, which has the highest HIV prevalence in South Africa and low practice of traditional circumcision. TMC coverage decreased in that time, due to the replacement of TMCs with MMC-Ts and men circumcised through VMMC interventions before or instead of through TMIC. Furthermore, while circumcision coverage is still increasing, over 5 million men aged 15-49 years remain uncircumcised and there are significant gaps to reaching 80\% coverage among men aged 15-49 years in many districts. However, 12 districts achieved greater than 80\% coverage of medical circumcision in the 15-24 years age group among whom intervention circumcisions focused, and continuing to provide VMMC services on this scale as well as focussing programmes in specific areas will ensure many more districts will achieve these targets in the near future. The number of MMCs conducted were largely similar in models with and without the inclusion of programme data, however, we found considerable differences in KwaZulu-Natal, which has the largest HIV prevalence and largest uptake in MMC services. It is therefore important to consider both data sources when estimating MC coverage. 

The granular results produced can be used to identify areas and key age groups in which either circumcision coverage is low or there are large numbers of uncircumcised men. Furthermore, stratifying by circumcision type allows VMMC programmes to make programmatic decisions about how to intervene in areas where TMCs remain high or where TMC is most prevalent. For HIV prevention purposes, it may be beneficial to offer "recircumcision" if TMCs are commonly only partial circumcision that does not involve complete removal of the foreskin. Furthermore, when assessing the impact of the VMMC programmes on HIV incidence, our estimates stratified by circumcision type enable models to account for different protective effects of full medical circumcision versus traditional circumcision.

There were several circumcision dynamics which are anecdotally known, but about which limited data were available and which required strong modelling assumptions. The first, was surrounding the proportion of traditional circumcisions conducted using medical methods (MMC-T) as ensuring safe and complete circumcisions for young men has been a major initiative for the VMMC programme. We made prior assumptions about the proportion of circumcisions in TMIC settings that were medical in recent years based on expert knowledge of programme managers and evidence of years in which large number of MMCs were reported in districts that have high levels of traditional circumcision. In 2020, the national health information system began explicitly recording the number of MMC-T, so quantitative data on this will be available for future analyses. The second, was uncertainty was about re-circumcision of men who were previously traditionally circumcised. We assumed that rates of re-circumcision were sufficiently low that it was not important to incorporate into the model. Future household surveys in South Africa plan to separately capture data on medical and traditional circumcision, which will substantiate or guide revision of this assumption. However, re-circumcision of those who previously underwent only partial circumcision could be important for maximising the HIV prevention impact of VMMC.

The models presented here have several limitations and opportunities for further development. First, surveys typically record self-reported circumcision status, which may be susceptible to misreporting as a result of social desirability bias. Many cultures that promote male circumcision as a rite of passage into manhood and some studies using physical examinations have shown that there is some misreporting in circumcision status \citep{lagarde2003acceptability, lissouba2011adult}. Second, the classification of circumcision type is based on self-reported questions from surveys, which may be subject to similar misreporting errors due to confusion between medical and traditional circumcision practices. Third, a complete-case analysis was used, that did not adjust for non-response bias in the self-reported circumcision status. Fourth, the models used did not make use of any covariates that are predictive of circumcision coverage, such as ethnicity, culture, or religion, to improve precision of the estimates. Extending the models in this way could allow for further targeting of VMMC services for specific population groups. Fifth, the model did not account for any uncertainty about the male population sizes by age and district. Small area population estimates are uncertain in many countries where the most recent census was long ago (in South Africa the last census was in 2011). Finally, the model did not explicitly represent migration between districts and its impact on district probabilities of circumcision and the corresponding coverage over time. It is suspected that many young men in South Africa move away from rural homes for work and return for their traditional male initiation ceremony. This is particularly the case with Xhosa men from Eastern Cape who move to Western Cape for work. When compiling the programme dataset used in the case study in this paper, we redistributed MMC-T conducted in the Eastern Cape to Xhosa populations residing in the Western Cape and Gauteng under the assumption that this is the normal place of residence for some men circumcised in the Eastern Cape, but future work should more explicitly model these migration dynamics and how they have changed over age and time.


\section*{Acknowledgements}


This research has been supported by the American People and the President's Emergency Plan for AIDS Relief (PEPFAR) through USAID under the terms of Cooperative Agreement 72067419CA00004 to HE$^2$RO and Imperial College London and through the Centers for Disease Control and Prevention (CDC), the Bill and Melinda Gates Foundation (INV-019496, INV-006733), National Institute of Allergy and Infectious Disease of the National Institutes of Health under award number R01AI136664, and the MRC Centre for Global Infectious Disease Analysis (reference MR/R015600/1), jointly funded by the UK Medical Research Council (MRC) and the UK Foreign, Commonwealth \& Development Office (FCDO), under the MRC/FCDO Concordat agreement and is also part of the EDCTP2 programme supported by the European Union. 

The findings and conclusions in this manuscript are those of the authors and do not necessarily represent the official position of the funding agencies.

\newpage 


\bibliographystyle{mattnat}
\bibliography{Refs}

\newpage 


\begin{landscape}
{\linespread{1}
\newpage
\footnotesize 

\begin{longtable}[c]{llc cc ccc}
	\hline
	&	& \multicolumn{3}{c}{\textbf{Coverage}} & \textbf{Change since 2008}& \multicolumn{2}{c}{\textbf{Population (in 1000s)}}  \\[1pt] 
	\cmidrule(lr){3-5}
	\cmidrule(lr){6-6}
	\cmidrule(lr){7-8}
	\multicolumn{2}{l}{\textbf{Region}} & \textbf{Total}  & \textbf{Medical} & \textbf{Traditional} & \textbf{Total} &  \textbf{Circumcised} & \textbf{Uncircumcised} \\[3pt] 
	\hline
	\vspace{-5pt}
	\endhead
	\\[-5pt]\hline 
	\caption{Estimated Total, Medical (MMC) and Traditional (TMC) circumcision coverage for 2020 among men aged 15-49, along with the absolute change in total MC coverage from 2008 to 2020 and the number of circumcised and uncircumcised men aged 15-49 in 2020.}
	\endfoot
	\multicolumn{2}{l}{\textbf{South Africa}} & 65.7\% (64.2\%--69.5\%) & 44.0\% (42.4\%--47.8\%) & 21.6\% (20.9\%--22.3\%) & 22.2\% (20.8\%--26.0\%) & 10398 (10165--11003) & 5432 (4827--5666) \\[5pt]
	\multicolumn{2}{l}{\textbf{Eastern Cape}}  & 74.2\% (72.5\%--78.3\%) & 25.3\% (23.3\%--29.5\%) & 48.9\% (47.6\%--50.2\%) & 9.9\% (8.3\%--13.4\%) & 1133 (1108--1196) & 395 (332--420) \\
		& A Nzo & 70.0\% (64.3\%--77.0\%) & 37.2\% (33.7\%--44.3\%) & 32.9\% (28.1\%--38.0\%) & 19.3\% (14.5\%--27.7\%) & 106 (97--116) & 45 (35--54) \\ 
		& Amathole & 75.9\% (73.0\%--79.4\%) & 12.0\% (10.1\%--15.6\%) & 63.9\% (61.0\%--66.3\%) & 7.9\% (6.6\%--11.3\%) & 135 (130--141) & 43 (37--48) \\ 
		& Buffalo City & 84.3\% (78.8\%--88.8\%) & 19.7\% (15.1\%--27.2\%) & 64.6\% (59.7\%--68.0\%) & -0.3\% (-4.2\%--5.1\%) & 168 (157--177) & 31 (22--42) \\ 
		& C Hani & 67.2\% (64.1\%--70.5\%) & 12.1\% (10.1\%--15.1\%) & 55.2\% (52.5\%--57.7\%) & 7.8\% (6.6\%--10.5\%) & 105 (100--110) & 51 (46--56) \\ 
		& Joe Gqabi & 73.5\% (68.4\%--78.1\%) & 15.2\% (12.8\%--18.4\%) & 58.4\% (52.7\%--63.7\%) & 5.7\% (4.2\%--9.0\%) & 57 (53--61) & 21 (17--25) \\ 
		& N Mandela Bay & 75.1\% (71.0\%--82.9\%) & 34.0\% (29.6\%--41.3\%) & 41.1\% (38.6\%--44.7\%) & 12.9\% (10.2\%--20.0\%) & 222 (210--245) & 74 (51--86) \\ 
		& O Tambo & 76.8\% (73.4\%--83.9\%) & 30.8\% (28.2\%--37.3\%) & 46.0\% (42.9\%--48.8\%) & 12.3\% (10.0\%--18.1\%) & 265 (253--289) & 80 (56--92) \\ 
		& Sarah Baartman & 60.1\% (55.1\%--65.1\%) & 25.2\% (22.0\%--29.7\%) & 34.9\% (30.3\%--40.9\%) & 11.5\% (9.4\%--15.7\%) & 76 (69--82) & 50 (44--56) \\[5pt]
 	\multicolumn{2}{l}{\textbf{Free State}}  & 65.9\% (63.2\%--69.6\%) & 49.9\% (47.3\%--54.4\%) & 16.0\% (14.2\%--18.7\%) & 29.8\% (27.7\%--34.3\%) & 487 (467--514) & 252 (225--272) \\  
		& Fezile Dabi & 69.4\% (63.9\%--76.0\%) & 54.9\% (50.2\%--61.2\%) & 14.5\% (9.9\%--19.8\%) & 25.3\% (21.5\%--30.1\%) & 91 (84--100) & 40 (31--47) \\ 
		& Lejweleputswa & 66.4\% (61.5\%--71.3\%) & 51.2\% (48.0\%--56.0\%) & 15.2\% (12.0\%--20.1\%) & 34.3\% (31.6\%--38.7\%) & 112 (104--120) & 57 (48--65) \\ 
		& Mangaung & 62.7\% (58.0\%--67.6\%) & 45.8\% (42.0\%--51.1\%) & 16.9\% (13.6\%--20.7\%) & 25.3\% (22.4\%--32.2\%) & 140 (129--151) & 83 (72--94) \\ 
		& T Mofutsanyana & 68.2\% (64.0\%--72.6\%) & 52.3\% (49.6\%--56.4\%) & 16.0\% (12.5\%--19.7\%) & 36.1\% (33.3\%--40.3\%) & 126 (119--134) & 59 (51--67) \\ 
		& Xhariep & 57.9\% (50.8\%--64.9\%) & 37.5\% (33.4\%--44.1\%) & 20.3\% (15.1\%--27.8\%) & 18.6\% (14.3\%--24.6\%) & 18 (16--20) & 13 (11--15) \\[5pt]
	\multicolumn{2}{l}{\textbf{Gauteng}}  & 67.7\% (65.5\%--72.3\%) & 48.3\% (46.0\%--53.2\%) & 19.3\% (17.9\%--21.3\%) & 19.4\% (17.9\%--24.0\%) & 3202 (3101--3423) & 1530 (1309--1632) \\ 
		& Ekurhuleni & 67.6\% (64.0\%--72.5\%) & 41.9\% (38.2\%--46.2\%) & 25.8\% (22.1\%--30.0\%) & 18.9\% (17.2\%--22.8\%) & 835 (790--895) & 399 (340--445) \\ 
		& Johannesburg & 69.4\% (66.1\%--75.1\%) & 55.5\% (51.8\%--61.3\%) & 13.9\% (11.4\%--16.3\%) & 21.5\% (19.5\%--27.0\%) & 1256 (1196--1360) & 554 (451--614) \\ 
		& Sedibeng & 72.9\% (68.5\%--78.7\%) & 53.6\% (49.8\%--59.9\%) & 19.2\% (16.1\%--23.3\%) & 17.9\% (15.5\%--23.1\%) & 206 (194--223) & 77 (60--89) \\ 
		& Tshwane & 63.6\% (59.3\%--68.3\%) & 42.9\% (39.2\%--47.6\%) & 20.7\% (17.7\%--24.8\%) & 16.8\% (15.2\%--20.8\%) & 715 (666--768) & 409 (356--457) \\ 
		& West Rand & 67.7\% (62.9\%--73.5\%) & 47.3\% (42.7\%--53.0\%) & 20.4\% (16.6\%--25.5\%) & 21.1\% (18.7\%--24.5\%) & 190 (177--207) & 91 (74--104) \\[5pt]
	\multicolumn{2}{l}{\textbf{KwaZulu-Natal}} & 61.5\% (58.8\%--68.3\%) & 56.1\% (53.4\%--62.8\%) & 5.3\% (4.6\%--6.0\%) & 40.5\% (38.2\%--46.8\%) & 1810 (1731--2011) & 1135 (935--1215) \\
		& Amajuba & 60.0\% (54.6\%--68.0\%) & 54.7\% (50.4\%--61.6\%) & 5.3\% (2.5\%--8.8\%) & 39.1\% (34.9\%--45.9\%) & 87 (79--98) & 58 (46--66) \\ 
		& eThekwini & 58.2\% (54.1\%--68.7\%) & 53.2\% (49.0\%--63.5\%) & 5.1\% (4.2\%--6.0\%) & 37.3\% (34.4\%--46.9\%) & 669 (622--790) & 480 (360--528) \\ 
		& Harry Gwala & 64.8\% (57.8\%--73.2\%) & 51.1\% (46.5\%--59.3\%) & 13.7\% (8.4\%--21.1\%) & 39.9\% (35.4\%--45.7\%) & 74 (66--83) & 40 (30--48) \\ 
		& iLembe & 52.8\% (49.3\%--60.2\%) & 48.5\% (44.9\%--55.9\%) & 4.3\% (2.6\%--6.2\%) & 30.5\% (27.6\%--36.9\%) & 89 (83--101) & 79 (67--85) \\ 
		& King Cetshwayo & 66.0\% (62.1\%--73.1\%) & 62.4\% (58.8\%--69.8\%) & 3.7\% (2.4\%--5.9\%) & 42.5\% (38.6\%--51.3\%) & 151 (142--167) & 78 (61--87) \\ 
		& Ugu & 63.6\% (59.3\%--73.5\%) & 56.1\% (52.0\%--65.7\%) & 7.5\% (4.8\%--11.2\%) & 45.5\% (41.0\%--53.8\%) & 130 (122--151) & 75 (54--83) \\ 
		& uMgungundlovu & 70.1\% (65.7\%--75.6\%) & 65.7\% (61.5\%--71.5\%) & 4.4\% (2.6\%--6.9\%) & 49.7\% (46.6\%--54.9\%) & 214 (201--231) & 91 (74--105) \\ 
		& Umkhanyakude & 61.8\% (55.8\%--69.6\%) & 55.8\% (50.2\%--63.9\%) & 6.0\% (2.9\%--11.6\%) & 44.5\% (40.2\%--53.0\%) & 92 (83--103) & 57 (45--66) \\ 
		& Umzinyathi & 71.8\% (68.5\%--80.2\%) & 67.0\% (63.9\%--74.6\%) & 4.8\% (3.4\%--6.9\%) & 48.3\% (44.9\%--55.1\%) & 88 (84--98) & 34 (24--39) \\ 
		& Uthukela & 68.8\% (65.1\%--75.5\%) & 64.7\% (61.6\%--71.9\%) & 4.1\% (2.7\%--6.0\%) & 45.8\% (42.1\%--52.4\%) & 115 (109--126) & 52 (41--58) \\ 
		& Zululand & 52.8\% (48.1\%--60.6\%) & 47.7\% (43.8\%--56.3\%) & 5.2\% (2.8\%--7.9\%) & 35.9\% (32.1\%--46.4\%) & 101 (92--116) & 91 (76--100) \\[5pt]	
 	\multicolumn{2}{l}{\textbf{Limpopo}} & 89.6\% (88.0\%--91.3\%) & 46.2\% (44.4\%--48.8\%) & 43.4\% (41.5\%--45.5\%) & 15.0\% (14.0\%--16.5\%) & 1199 (1178--1222) & 139 (116--160) \\
		& Capricorn & 89.1\% (84.9\%--92.3\%) & 58.8\% (55.3\%--63.1\%) & 30.4\% (26.3\%--34.9\%) & 18.4\% (16.3\%--20.9\%) & 255 (243--263) & 31 (22--43) \\ 
		& Mopani & 92.3\% (88.8\%--95.2\%) & 47.7\% (44.0\%--52.2\%) & 44.6\% (40.0\%--49.3\%) & 16.3\% (13.6\%--18.7\%) & 244 (235--252) & 21 (13--30) \\ 
		& Sekhukhune & 92.2\% (89.2\%--94.5\%) & 31.0\% (28.2\%--33.9\%) & 61.2\% (57.7\%--64.4\%) & 11.6\% (10.1\%--13.5\%) & 257 (249--264) & 22 (15--30) \\ 
		& Vhembe & 94.4\% (90.2\%--96.7\%) & 46.3\% (41.6\%--50.9\%) & 48.1\% (42.5\%--54.0\%) & 8.1\% (6.6\%--10.4\%) & 301 (288--308) & 18 (10--31) \\ 
		& Waterberg & 74.8\% (70.1\%--79.6\%) & 47.7\% (44.3\%--51.5\%) & 27.1\% (22.2\%--33.1\%) & 23.4\% (21.4\%--26.5\%) & 142 (133--151) & 48 (39--57) \\[5pt]
	\multicolumn{2}{l}{\textbf{Mpumalanga}} & 73.1\% (71.0\%--77.2\%) & 53.8\% (51.9\%--57.5\%) & 19.3\% (18.0\%--20.8\%) & 28.9\% (27.2\%--33.1\%) & 929 (902--980) & 341 (290--368) \\
		& Ehlanzeni & 69.2\% (66.2\%--75.0\%) & 51.6\% (48.8\%--56.5\%) & 17.6\% (15.0\%--20.0\%) & 26.8\% (24.4\%--32.0\%) & 311 (297--336) & 138 (112--152) \\ 
		& G Sibande & 59.4\% (55.6\%--64.2\%) & 48.2\% (45.6\%--52.8\%) & 11.1\% (9.3\%--13.0\%) & 24.3\% (22.1\%--28.2\%) & 204 (192--221) & 140 (123--153) \\ 
		& Nkangala & 86.8\% (83.9\%--89.7\%) & 59.9\% (57.9\%--64.1\%) & 26.9\% (24.3\%--29.5\%) & 33.4\% (30.8\%--36.5\%) & 414 (401--428) & 63 (49--77) \\[5pt]
	\multicolumn{2}{l}{\textbf{Northern Cape}} & 42.3\% (39.0\%--47.5\%) & 30.2\% (27.7\%--34.9\%) & 12.1\% (10.2\%--14.3\%) & 17.7\% (15.3\%--22.9\%) & 120 (111--135) & 164 (149--173) \\ 
		& Frances Baard & 49.5\% (45.1\%--55.5\%) & 41.3\% (38.0\%--48.2\%) & 8.2\% (5.7\%--11.2\%) & 24.4\% (21.2\%--32.5\%) & 45 (41--50) & 46 (40--50) \\ 
		& J T Gaetsewe & 59.9\% (53.4\%--68.6\%) & 36.8\% (31.7\%--44.9\%) & 23.1\% (15.9\%--32.3\%) & 22.6\% (18.6\%--30.1\%) & 35 (31--40) & 23 (18--27) \\ 
		& Namakwa & 19.1\% (15.2\%--24.6\%) & 13.2\% (10.4\%--17.5\%) & 6.0\% (3.0\%--10.6\%) & 5.8\% (3.9\%--9.5\%) & 5 (4--6) & 20 (18--21) \\ 
		& Pixley ka Seme & 35.9\% (31.5\%--42.1\%) & 22.1\% (19.4\%--28.7\%) & 13.8\% (9.9\%--18.7\%) & 10.1\% (7.4\%--17.0\%) & 16 (14--19) & 28 (25--30) \\ 
		& ZF Mgcawu & 29.7\% (26.4\%--35.4\%) & 20.8\% (18.2\%--26.3\%) & 9.0\% (6.7\%--11.9\%) & 12.2\% (9.6\%--17.4\%) & 20 (18--24) & 47 (43--49) \\[5pt]
	\multicolumn{2}{l}{\textbf{North West}} & 58.2\% (55.6\%--61.3\%) & 40.4\% (38.1\%--44.0\%) & 17.8\% (15.7\%--19.8\%) & 23.9\% (22.0\%--27.2\%) & 651 (622--686) & 468 (434--497) \\
		& Bojanala Platinum & 58.1\% (54.7\%--62.2\%) & 37.1\% (34.7\%--40.5\%) & 21.0\% (18.3\%--23.9\%) & 18.8\% (16.8\%--22.5\%) & 336 (316--360) & 242 (219--262) \\ 
		& Dr K Kaunda & 61.8\% (57.3\%--66.4\%) & 47.5\% (43.2\%--52.5\%) & 14.3\% (10.7\%--18.5\%) & 28.9\% (26.3\%--32.6\%) & 128 (119--138) & 79 (70--89) \\ 
		& Ngaka Modiri Molema & 52.9\% (47.3\%--60.6\%) & 40.7\% (36.5\%--48.9\%) & 12.2\% (8.6\%--16.7\%) & 27.1\% (23.2\%--34.2\%) & 120 (108--138) & 107 (90--120) \\ 
		& Ruth Segomotsi Mompati & 63.1\% (55.9\%--72.1\%) & 44.3\% (39.3\%--52.9\%) & 18.8\% (12.9\%--26.5\%) & 30.7\% (25.7\%--40.0\%) & 67 (59--76) & 39 (29--46) \\[5pt]
	\multicolumn{2}{l}{\textbf{Western Cape}} & 46.2\% (43.8\%--50.9\%) & 23.3\% (20.9\%--26.3\%) & 23.0\% (20.9\%--25.2\%) & 6.5\% (5.2\%--9.5\%) & 866 (821--954) & 1008 (920--1054) \\
		& Cape Town & 50.0\% (47.1\%--54.9\%) & 24.5\% (21.9\%--27.6\%) & 25.5\% (22.9\%--28.2\%) & 4.6\% (3.2\%--7.8\%) & 623 (587--684) & 623 (562--659) \\ 
		& Cape Winelands & 35.9\% (29.4\%--43.9\%) & 19.6\% (15.2\%--25.7\%) & 16.3\% (12.0\%--23.1\%) & 8.2\% (5.9\%--12.4\%) & 93 (76--113) & 165 (145--182) \\ 
		& Central Karoo & 56.8\% (42.5\%--71.1\%) & 24.1\% (18.5\%--30.8\%) & 32.7\% (19.8\%--49.9\%) & 12.0\% (8.4\%--16.7\%) & 10 (7--12) & 7 (5--10) \\ 
		& Garden Route & 42.8\% (36.0\%--50.5\%) & 25.2\% (21.3\%--31.2\%) & 17.6\% (12.3\%--23.3\%) & 13.3\% (10.9\%--17.5\%) & 63 (53--75) & 85 (74--95) \\ 
		& Overberg & 42.9\% (34.9\%--52.7\%) & 19.6\% (15.4\%--25.3\%) & 23.3\% (15.2\%--33.3\%) & 11.4\% (9.3\%--16.2\%) & 35 (28--43) & 46 (38--53) \\ 
		& West Coast & 34.6\% (29.4\%--41.6\%) & 18.8\% (14.5\%--26.4\%) & 15.8\% (11.7\%--21.8\%) & 8.8\% (5.4\%--14.6\%) & 43 (37--52) & 81 (73--88) \\ 
		\label{tab:results}
\end{longtable}}
\end{landscape}


\end{document}